\documentclass[12pt]{article}

\usepackage{amssymb}
\usepackage{amsmath}
\usepackage{amssymb}
\numberwithin{equation}{section}
\begin{document}
\begin{center}
{\Large\bf \boldmath  Irreversible Evolution of Open Systems 
and the  Nonequilibrium Statistical Operator Method}
 
\vspace*{6mm}
{A. L. Kuzemsky}   \\      
{\small   Bogoliubov Laboratory of Theoretical Physics,}\\
{\small    Joint Institute for Nuclear Research, 141980 Dubna, Moscow Region, Russia.\\
kuzemsky@theor.jinr.ru; \,  http://theor.jinr.ru/\symbol{126}kuzemsky}         
\end{center}

\vspace*{2mm}

\begin{abstract}
The effective approach to the foundation of the nonequilibrium statistical mechanics on the 
basis of dynamics was formulated by Bogoliubov in his seminal works. 
His ideas of reduced description were proved as very powerful and found
a broad applicability to quite general time-dependent problems of physics and mechanics. 
In this paper we  analyzed   thoroughly  the time evolution of  open systems in  context 
of the nonequilibrium statistical operator method (NSO). This method extends the statistical 
method of Gibbs to irreversible processes and incorporates the ideas of reduced description. 
The purpose of the present study was to elucidate the 
basic aspects of  the NSO method  and some few selected 
approaches to the nonequilibrium statistical mechanics. The suitable procedure of 
averaging (smoothing) and the notion of irreversibility were discussed in this context.
We were focused on the physical consistency of the  method as well as on its operational ability   
to emphasize and address a few important reasons for such  a workability.
%

%
%
\vspace*{2mm}
\end{abstract}
%
%
\newpage
\tableofcontents
\newpage
%
%
%
\section{Introduction}
%
%
%
\begin{flushright}
{\em As soon as one starts to think about \\the nonequilibrium phenomena the questions multiply.\\ 
G. Uhlenbeck}
\end{flushright}
The aim of statistical mechanics is to give a consistent
formalism for a microscopic description of   macroscopic behavior of matter in bulk.
Statistical mechanics formulates the consistent approach that successfully
describes the stationary macroscopic behavior of many-particle systems as fluids,
gases, and solids. It clarifies  also the thermodynamic concepts such as heat, temperature,
and entropy  from the underlying microscopic laws.\\
It is important to emphasize that in the structure of thermodynamics the one of its basic law, namely the
second law, differs very much from other general laws of physics. It is not an equation, but instead states
an inequality, which becomes an equality only in the limiting case of a reversible process. There are difficulties with
the realization of this limit, because a reversible process is one in which the thermodynamic system never 
deviates appreciable  from equilibrium. However, finite time process involves  a disturbance of equilibrium. As a result,
it is difficult (if not impossible) to derive the correct equations concerning time rates. It was said sometimes that
time appears in thermodynamics not as a quantity but only as the indicator  of the sense of a quantity, the change
of entropy. The goal of a theory is to describe the actual reality. Real systems are never close.
The nonequilibrium statistical thermodynamics aims to describe in the unifying manner irreversible phenomena
including nonequilibrium steady states and open systems.\\
The theoretical study of transport  processes in matter is a very broad and well explored field.
The methods of equilibrium and nonequilibrium statistical mechanics have been fruitfully applied to a large variety 
of phenomena and materials.
The central problem of nonequilibrium statistical mechanics is to derive a set of equations which
describe irreversible processes from the reversible equations of motion.
The consistent calculation of transport coefficients
is of particular interest because  one can get information on the microscopic
structure of the condensed matter. There exist a lot of theoretical methods for  calculation
of transport coefficients as a rule having a fairly restricted range of validity and
applicability.  The most extensively developed theory of transport processes is that based on the
Boltzmann equation. However, this approach has strong restrictions and can reasonably be applied 
to a strongly rarefied gas of point particles.\\
During the last decades, a number of schemes have been concerned with a more general and consistent approach to kinetic and transport 
theory~\cite{bogol46,bog62,nnb78,dpetr09,zub74,macl89,eu98,zwa65,zw70,rbal97,zwan01,galla14,kozlov08,vede11}.
In what follows, we discuss in terse form  that direction in the nonequilibrium statistical mechanics which is based 
on nonequilibrium ensemble formalism and
compare it briefly with some other approaches for  description of irreversible processes.\\
These approaches, each in its own way, lead us to substantial advances in the understanding of the
nonequilibrium behavior of many-particle classical and quantum systems. 
This field is very active and there are many aspects to the problem~\cite{kuzem17,kuzem07,kuzem05,kuzem11}.  
We survey concisely a formulation of the method of nonequilibrium
statistical operator, introduced  by  Zubarev~\cite{zub74} and some  of its 
applications to concrete problems, with the purpose of making these ideas easier for understanding and  applications.
The relation to other work is touched on briefly, but the  nonequilibrium statistical operator (NSO) method is considered 
of dominant importance. 
%
%
%
%
\section{The Ensemble Method} 
%
%
The formalism of equilibrium statistical mechanics has been developed by Gibbs~\cite{gib1}
to describe the properties of many-particle complex  physical systems. 
Central problem in the statistical physics of matter is that of accounting for the observed 
equilibrium and nonequilibrium properties of fluids and solids from a specification of the 
component molecular species, knowledge of how the constituent molecules interact,
and the nature of their surrounding.
It was  shown  rigorously by Kozlov~\cite{kozlov00,koz02} that  the Gibbs canonical distribution 
(Gibbs ensemble) is the only \emph{universal}  one whose density depends on energy and that 
is compatible with the axioms of thermodynamics.\\
The ensemble method, as it was formulated by Gibbs~\cite{gib1,mehra98}, have the great generality and the broad applicability to the equilibrium
statistical mechanics. The Gibbsian concepts and methods are used today in a number of different 
fields~\cite{kuzem17,kuzem07}. Ensembles are a far more
satisfactory starting point than \emph{assemblies}, particularly in treating time dependent 
systems~\cite{zub74,kozlov08,kuzem07}. 
An ensemble is characterized by the distribution (partition) function $f(\mathbf{p},\mathbf{q})$ which should satisfy Liouville 
equation 
\begin{equation}
  \frac{d f}{d t} = 0.
\end{equation}
This purely dynamical requirement reflects the fact that the points in the phase space $(\mathbf{p},\mathbf{q})$ representing the states of the system in an ensemble
do not interact. It is important to realize that the Liouville equation is an expression of the preservation of volumes of phase space.\\
Equilibrium ensemble theories are rooted in the fundamental principle of equal probability for the microstates of 
isolated systems~\cite{kuzem17,kuzem14}.
This principle (or postulate) is, in essence, a kind of statistical approximation but mechanical origin.\\
The notions of the Gibbs state and \emph{Gibbs distribution}~\cite{minl}, which play  an important role in determining equilibrium properties of statistical ensembles, 
were clarified substantially in the last decades. A \emph{Gibbs state} in probability theory and statistical mechanics is 
an \emph{equilibrium probability distribution} which remains invariant under future evolution of the system.\\ 
For systems in the state of statistical equilibrium, there
is the Gibbs distribution~\cite{gib1} by means of which it is possible to calculate an average value 
of any dynamical quantity. No such universal distribution has been formulated for irreversible
processes. Thus, to proceed to the solution of problems of statistical mechanics of
nonequilibrium systems, it is necessary to resort to various approximate methods. 
In addition, the   Gibbs distributions, have a non-trivial common property: subject to 
certain constraints it maximize a functional known in statistical mechanics as entropy, 
and in information theory, probability theory and mathematical statistics as information. 
The approach based on the information theory in the spirit of the principle of maximum entropy
has been used in numerous works on statistical mechanics~\cite{zub74,kuzem17,jayn03,koz06,kuzem16}
to derive the fundamental statistical mechanical distributions. \\
Kubo~\cite{kuzem17,kuzem11,kuzrnc18} derived the quantum statistical expressions for transport coefficients such as electric  
conductivity. This approach considered the case of mechanical disturbances such as an electric field.
The mechanical disturbance was expressed as a definite perturbing Hamiltonian and the deviation 
from equilibrium caused by it can be obtained by perturbation theory. On the other hand, thermal
disturbances such as density and temperature gradients cannot be expressed as a perturbing Hamiltonian
in an unambiguous way~\cite{zub74,kuzem17,kuzem07}.  
It is worth noting that there exist  a substantial distinction of the  standard linear 
response theory and of  the Zubarev's method of the nonequilibrium statistical 
operator~\cite{zub74,kuzem17,kuzem07}. In essence, the linear response theory is an expansion
from the \emph{global} equilibrium state whereas the nonequilibrium statistical operator approach 
uses the expansion from the \emph{local} (quasi-equilibrium) state. Hence it may provide a more 
consistent description of various nonequilibrium nonlinear processes. \\ 
Temporal evolution of complex statistical systems  represents in itself the fundamental aspects 
of the nonequilibrium statistical 
physics~\cite{bogol46,bog62,nnb78,dpetr09,zub74,macl89,eu98,kuzem17,kuzem07,kuzrnc18}. 
It includes also the stochastic nonequilibrium processes~\cite{nnb78,ross08,scha14,tome15}. A stochastic 
process may be considered as the description of a random phenomenon evolving in time that is 
governed by certain laws of probability. To describe these complicated 
behaviors   various generalized statistical mechanics methods for complex systems were 
developed~\cite{bogol46,bog62,nnb78,dpetr09,zub74,macl89,eu98,kuzem17,kuzem07,kuzrnc18,ross08,scha14,tome15}.
The answers on a question of how to obtain an irreversible description of temporal processes are 
numerous~\cite{zub74,macl89,eu98,kuzem17,demi14}. 
Usually the statistical-mechanical theory of transport is divided into two related problems: 
the mechanism of the approach to equilibrium, and the
representation of the microscopic properties in terms of the macroscopic fluxes.\\ 
It is well known~\cite{zub74} that in a standard thermodynamic approach one deals with only a small number of state variables
to determine the properties of a uniform equilibrium system. To deal with irreversible processes in systems not too far  from
equilibrium, one divides the system into small subsystems and assumes that each subsystem is in \emph{local equilibrium }~\cite{zub74,kuzem17,kuzem07}, i.e., it can
be treated as an individual thermodynamic system characterized by the small number of physical variables. For continuous systems, there is
a temperature $T$ associated with each subsystem, and $T \sigma$ is the \emph{dissipation}; here $\sigma$ is the "entropy production", which is
defined as the time rate of the entropy created internally by an irreversible process~\cite{msuz11,msuz11a,msuz12}. 
Note, that according to the law of thermodynamics, reversible evolution is an evolution with constant entropy.\\
If $T \sigma$  is calculated for various irreversible processes, it is always found to have the form
 $$T \sigma = \sum_{i}J_{i}X_{i} > 0,$$
where $J_{i}$ are flows of the matter, heat, etc., and $X_{i}$ are generalized driving forces for vector transport processes or for chemical reactions, etc.
The $J_{i}$ and $X_{i}$ are linearly related when the system is not too far from equilibrium. Thus
$$J_{i} = \sum_{i}L_{ij}X_{j},$$
where the $L_{ij}$ are called phenomenological transport coefficients. The aim of the nonequilibrium statistical mechanics 
is to calculate these transport coefficients microscopically.\\
Great efforts have been directed by numerous authors toward establishing the theory of irreversible processes on a microscopic
basis~\cite{zub74,kuzem17,kuzem07}.
Zwanzig~\cite{zwa60}   have reformulated the methods of Prigogine and
Van Hove. His reformulation was characterized by extensive use of Gibbsian ensembles. Projection operators in the space of all possible
ensemble densities were used to separate an ensemble density into a \emph{relevant} part, required for the calculation of ensemble
averages of specified quantities, and the remaining \emph{irrelevant} part. This was a generalization of the common separation of a density
matrix into diagonal and nondiagonal parts, as used in  derivation of the master equation.\\
Zwanzig showed that the Liouville equation is the natural starting point for a theory of time-dependent processes in statistical mechanics. He
considered the ensemble density (phase space distribution function or density matrix)  $f(t)$ at time $t$. The average of a dynamical variable $A$
(function, matrix or operator) at time $t$ is $\langle A;f(t) \rangle$.  
Thus, the main tool in Zwanzig reformulation~\cite{zwa65,zw70,zwan01,zwa60} was the use of projection operators in the Hilbert space of Gibbsian ensemble densities. Projection
operators are a convenient tool for the separation an ensemble density into a \emph{relevant} part, needed for the calculation of mean values of
specified observables, and the remaining \emph{irrelevant} part. The \emph{relevant} part was shown to satisfy a kinetic equation which
is a generalization of Van Hove master equation; diagram summation methods were not used in this approach.\\
Hence, the Gibbs ensemble in statistical mechanics serves as a microscopic formulation of equilibrium thermodynamics,
and the fluctuation-dissipation theorem provides a microscopic connection to the system response functions and
transport coefficients which characterize small departures from equilibrium. Far from equilibrium, Lyapunov expansion
is a property with the potential to provide a useful microscopic description, when local definitions of quasi-equilibrium
quantities, such as temperature and pressure, may no longer have meaning. The Lyapunov exponent measures the rate at which a system "forgets" its initial
conditions. The transport coefficients are those response functions of the system that also measure a "forgetting". 
 For example, scattering erases a particle's  memory of its original velocity and so give rise to a finite diffusion coefficient.
Many authors have been exploring the connection between transport coefficients and Lyapunov 
exponents~\cite{kuzem17,kuzem07}.
%
%
%
%
\section{Bogoliubov's Fundamental Results} 
%
%
%
Bogoliubov contributed greatly to equilibrium and nonequilibrium statistical mechanics and 
received many fundamental results. 
The intense current interest in the statistical mechanics of irreversibility  is in the 
foundation of the nonequilibrium statistical mechanics on the basis of dynamics.
The important contribution of Bogoliubov are his notable papers 
on dynamical theory and physical kinetics~\cite{bogol46,bog62,nnb78}. There he introduced the key concept of the hierarchy 
of relaxation times in statistical physics. This method relies substantially on the existence of 
the natural fine-scale mixing occurring in  dynamics. \\
Bogoliubov method emphasized three important points:
(i) Temporal Evolution; (ii) Reduced Description (Relevant Variables);(iii) Averaging Procedure.
These ideas have a fundamental role in the description of multi-scale systems characterized by 
hierarchical configurations of complex many-particle systems.\\
Bogoliubov approach leads to a systematic development of the equations describing the time 
evolution of the two-particle, three-particle, etc., distribution functions, and treats
the time development as occurring in rather well-defined stages. At each stage the system 
has "\emph{forgotten}" more and more of the information contained
in the initial $n$-particle distribution function. Thus Bogoliubov assumed that after a time 
of the order of the duration of a collision between
molecules, all the higher-order distribution functions will depend on time only as functionals 
of the single-particle distribution function. At the next,
or "hydrodynamic" stage, only the first few moments of the single-particle distribution 
function, i.e., the local values of density, temperature, and
flow velocities, are needed to describe the evolution of the system. \\
A method of obtaining a system of coupled equations 
for the probability densities for groups of one or more particles was proposed. This has proved 
to be the most effective method in statistical mechanics for equilibrium and non-equilibrium 
systems to date. In his approach  Bogoliubov clarified how stochastic behavior, which is specific 
for a macroscopic description, arises in a purely mechanistic approach, in which microscopic 
equations of dynamical theory are used.\\ 
Bogoliubov introduced the concept of stages of the evolution - chaotic, kinetic, and hydrodynamic 
and the notion of the time scales, namely, interaction time $\tau_{\textrm{coll}}$, 
free path time $\tau_{\textrm{fp}}$, and time of macroscopic relaxation $\tau_{\textrm{rel}}$, 
which characterize these stages, respectively. At the chaotic stage, 
the particles synchronize, and the system passes to a \emph{local equilibrium}. At the kinetic 
stage, all distribution functions begin to depend on time via the one-particle
function. Finally, at the hydrodynamic stage, the distribution functions depend on time 
via macroscopic variables, and the system approaches equilibrium.\\
For dilute gas, when the time of intermolecular interaction (interaction
time, $\tau_{\textrm{coll}}$) is less than the free path time, there are two kinds of processes: \emph{fast}
and \emph{slow}. Hence for the fast processes characteristic time is  of the order 
$\tau_{\textrm{coll}}$ and for slow processes - of the order $\tau_{\textrm{fp}}$.\\
Moreover, Bogoliubov for the first time formulated the \emph{boundary conditions} for the 
chain of the $n$-particle distribution functions. These conditions correspond to the
\emph{reduced description} and replace the Boltzmann's hypothesis of the \emph{molecular chaos}.
Indeed, in the Boltzmann approach the irreversibility feature was connected with the assumption 
that information about individual molecular dynamics is forgotten
after the collisions (for rarefied gas). In contrast, only the probability distribution of 
velocities among the particles was remembered. Hence, this lack of memory
(or continual randomization) namely may be considered as the real source of irreversibility. 
As it was showed clearly by Bogoliubov, the system should be \emph{large enough} in order to the 
randomization assumption was reasonably applicable.\\
In this connection, a significant contribution in the rigorous treatment of the thermodynamic 
limit~\cite{kuzem16} was made by   Bogoliubov, who developed a general formalism for establishing 
of the limiting distribution functions in the form of formal series powers of the density.
Bogoliubov outlined the method of justification of the thermodynamic limit and derived the 
generalized Boltzmann equations from his formalism. Bogoliubov also introduced the important clustering principle. Bogoliubov conjectured that it is 
often convenient to separate the dependence on
momenta and consider distribution functions, which will depend only on coordinates passing 
then to the thermodynamic limit. Thus, on the basis of his equations
for distribution functions and the cluster property, the Boltzmann equation was first 
obtained without employing the molecular chaos hypothesis.
%
%
\section{Nonequilibrium Ensembles} 
%
%
Here, we    remind very briefly 
the main streams of the  nonequilibrium ensembles approaches to  statistical mechanics. 
Beginning have been made by Lebowitz and Bergmann~\cite{berg55,leb57,leb59,leb62} and some other authors.\\ 
The central statement of the statistical-mechanical picture is the fact that it is practically 
impossible to give a complete description of the state of a complex macroscopic system. We must substantially  reduce 
the number of variables and confine ourselves to the description of the system which is considerably
less then complete. The problem of predicting  probable behavior of a system at some specified 
time is a statistical one~\cite{bogol46,bog62,nnb78,zub74}. It is useful and
workable to employ the technique of representing the system by means of an ensemble~\cite{gib1} consisting of a 
large number of identical copies of a single system under consideration. The state of  ensemble
is then described by a distribution 
function $\rho (\mathbf{r}_{1} \ldots \mathbf{r}_{n}, \mathbf{p}_{1} \ldots \mathbf{p}_{n}, t)$ in the phase space
of a single system. This distribution function is  chosen so   that averages over the ensemble are in exact 
agreement with the incomplete (macroscopic) knowledge of the state of the system at some specified time.
Then the expected development of the system at subsequent times is modelled via the average behavior of members of the
representative ensemble. It is evident that there are many different ways in which an ensemble could be
constructed. As a result, the basic notion, the distribution function $\rho$ is not uniquely defined.
Moreover, contrary to the description of a system in the state of thermodynamic equilibrium which is
only one for fixed values of volume, energy, particle number, etc., the number of 
nonequilibrium states is large. \\
The precise definition of the nonequilibrium state is quite difficult
and complicated task, because of this state is not  specified uniquely. However in certain cases the thorough considerations
lead to the establishment of various time scales~\cite{bogol46,bog62,nnb78,zub74}. In fact, the quasi-equilibrium and
nonequilibrium formulations are quite similar in structure and functional dependence. Thus,
a large and important class of transport processes can  reasonably be  modelled in terms of a reduced number of macroscopic
\emph{relevant variables}~\cite{zub74}.\\ 
This line of reasoning has led to  seminal ideas on the
construction of   Gibbs-type ensembles for  nonequilibrium systems. Such a program is essentially 
designed to develop a statistical-mechanical description
of nonequilibrium  processes, motivated by the success of the statistical mechanics of Gibbs for the equilibrium state. 
The possibility of carrying over Gibbs approach to nonequilibrium statistical mechanics was anticipated
by Callen and Welton~\cite{cawe} in connection with the fluctuation-dissipation theorem~\cite{zub74}.\\ In attempting to
develop such a theory it must be kept in mind that in order for a system to approach s steady state, or to remain 
in a nonequilibrium stationary state, it cannot be isolated but must be in contact with surroundings (reservoirs) 
which maintain gradients within it.\\
The attempt to construct general nonequilibrium ensembles was carried out by Lebowitz and Bergmann~\cite{berg55,leb57,leb59,leb62}.
They used a model reservoir that, as far as the system was concerned, always had the same appearance and consisted of an infinite number of
independent, identical components, each of which interacted with the system but once. Thus the process can be considered as the truly stationary.
They assumed also that there was an impulsive interaction between system and reservoir components. So, it was not necessary to deal with the total,
infinite, phase-space of the reservoir. In this approach these reservoirs played the role of thermodynamic temperature baths. The ensemble, representing
a system in contact with such reservoirs, obeys an integro-differential equation in $\Gamma$-space, containing both the Liouville equation and a
stochastic integral term that describes the collision with the reservoirs.  The Onsager relations~\cite{zub74,kuzem07} were obtained
without any reference to fluctuation theory and without the assumption of detailed balancing. 
They derived the stationary 
distribution (via an iteration procedure). Lebowitz~\cite{leb59} has found exact stationary nonequilibrium solutions for some simple systems, and 
has introduced  a simple relaxation-type method for finding approximate stationary solutions for the distribution function. 
This program was not fully successful. The difficult aspect of this
approach  was in the handling in detail the interaction between the system and reservoir, making its utility uneasy.\\
A method similar to the method of nonequilibrium statstical operator~\cite{zub74} was formulated by McLennan~\cite{macl89}. His method
is based on the introduction of external forces of a nonconservative nature, which describe the influence of the surroundings or of a thermal bath
on the given system. In other words, McLennan operates with the energy and particle reservoirs and movable pistons in contact with the system. The
evolution equation for the distribution function $f_{U}$ within the McLennan approach has the following form
\begin{equation}\label{e}
 \frac{\partial f}{\partial t} + [f,H]  + \frac{\partial f F_{\alpha}}{\partial p_{\alpha}}  = 0,
\end{equation}
where 
\begin{equation}\label{e}
 F_{\alpha}  = - \int g X \frac{\partial U}{\partial q_{\alpha}}d \Gamma_{s}.
\end{equation}
Here U is the Hamiltonian of the interaction with the surroundings, $f = \int f_{U}d\Gamma_{s},$ $g = \int f_{U}d\Gamma,$ $f_{U} = f g X,$
$q_{\alpha}$ and  $p_{\alpha}$ are the coordinates and momenta of the system. The quantity $F_{\alpha}$ has the meaning 
of a ``\emph{force}'' representing the action of the surrounding on the system (for details see Refs.~\cite{zub74,macl89,kuzem17}).\\ 
Nevertheless, it was become clear that the sophisticated ideas are required for describing
a situation when we need to average over reversible dynamical equations
and acquire the irreversible equations; the numerous approaches for solution of this problem 
were proposed. 
The essence of the task was summarized by Mayer~\cite{mayer61}:
``The properties of a macroscopic classical system consisting of some $10^{23}$ molecules are determined by a probability 
density function $W$ of the complete $\Gamma$ space of moments and coordinates of all the molecules. This probability 
density function is that of the ensemble representing the totality of all experimental systems prepared according to the 
macroscopic specifications. The entropy is always to be defined as the negative of $k_{B}$ times the integral 
over the distinguishable phase space of $W \textrm{ln} W$. However, the total probability density function $W$, even for a 
thermodynamically isolated system, does not obey the Liouville equation, $ \partial W/\partial t = L W$, since small 
fluctuations due to its contact with the rest of the universe necessarily ``smoothes'' $W$, by 
smoothing the direct many-body correlations in its logarithm. This smoothing is the cause of the entropy increase, 
and in systems near room temperature and above, in which there is heat conduction or chemical species diffusion, 
the smoothing keeps the true entropy numerically equal to that inferred from the local temperatures, pressures, and 
compositions. This, however, is by no means necessarily general. The criterion of thermodynamic 
isolation is not that the complete probability density function $W$ is unaffected by the surroundings, but that 
reduced probability density functions $w_{n}$ 
in the $\Gamma$ space of $n = 2,3,\ldots $ molecules evolve in time as if the system were unaffected by the surroundings. 
This criterion is sufficient to give a mathematically definable method of ``smoothing'' the complete probability density 
function. The smoothing consists of replacing the direct many-body correlations in $ \textrm{ln} W$ by their average 
n-body values, $n = 2,3,\ldots $, such that the smaller reduced probability density 
functions $w_{n}$ are unaffected''.\\
Method of averaging in complex many-particle (nonlinear) systems was elaborated by Bogoliubov, Krylov and Mitropolski. It  was 
discussed  and generalized by various authors~\cite{bmitr,mitr2,sam94,arnold}.
In this context, it was shown by Bogoliubov~\cite{bog82} that the mixing property arising in ergodic theory is not necessary for
statistical systems for \emph{any finite} volume and number of particles. Of importance is
only the appropriate behavior of the \emph{limiting average values} of the macroscopic
quantities at $t \rightarrow \infty$ after the transition to the limit of statistical mechanics~\cite{bpk69} has
been performed.\\ 
Bogoliubov emphasizes~\cite{bog82}  the fact that ergodic theory in its standard form is not
sufficiently established. In order to explain this idea, some model systems were
investigated, i.e. the problem of interaction of the particle with quantum field.
The open mathematical questions in this field were pointed and it was especially
stressed that one has not succeeded yet in rigorous proving the properties of
many-particle systems which were required by the basic postulate of statistical mechanics. These conclusions by 
Bogoliubov~\cite{bog82} anticipated the subsequent critical arguments~\cite{earm96} by  Earman and  Redei and other authors 
"why ergodic theory does not explain the success of equilibrium statistical mechanics". 
It is worth noting that relatively recently  a certain progress in this field 
was achieved in the remarkable publications~\cite{kozl03,bogach07}, where  generalizations of the 
classical Birkhoff and von Neumann ergodic theorems were presented.   The time average
with the aid of special summation methods was replaced by a more general average, including 
some density $\varrho$. The character of averaging  differs fundamentally
from classical uniform averaging.\\
The main theoretical aspects of the problem of averaging were
clarified by Bogoliubov in his  seminal works. He formulated
the proper method of "\emph{averaging}" and carried out a full mathematical justification of it.
This line of reasoning was refined and developed further by  Bogoliubov and Mitropolsky~\cite{bmitr,mitr2}.
They have elaborated perturbation methods to obtain asymptotic solutions without secular terms.
In this approach~\cite{bmitr,mitr2},
the main intention  was to  find a transformation of variables
which would separate the "\emph{slow}" variables  from the "\emph{fast}" ones.
Subsequently, Bogoliubov worked out a rigorous theory of the averaging method  and showed
that it is naturally related to the existence of a certain transformation of variables
that enables one to eliminate the time $t$ from the right-hand sides of the corresponding equations to within
an arbitrary accuracy relative to the small parameter $\varepsilon$. At the same time, invoking subtle
physical considerations, he showed how to construct not only the first-approximation system
(averaged system), but also averaged systems in higher approximations, whose solutions approximate
the solutions of the original (exact) system to within an arbitrary prescribed accuracy~\cite{bmitr,mitr2,sam94,arnold}.\\
A brilliant example of the physical problem where
separation of the "\emph{slow}" variables $a$ from the "\emph{fast}" ones $\psi$ within Krylov-Bogoliubov method was especially successful, was the work
of Bogoliubov and  Zubarev~\cite{dnz55} on plasma in the magnetic field. 
It was shown   later how this method can be generalized to higher order of the perturbation expansion in the case 
of nearly periodic or nearly quasi-periodic systems.\\
In the present context, it is of importance to stress that 
the averaging method in the nonlinear mechanics~\cite{bmitr}  has much in common with the statistical 
mechanics~\cite{zub74,arnold}. These ideas were implemented ingeniously by Zubarev~\cite{zub74} 
in his NSO method.
Indeed, there is a close analogy with the statistical mechanical problematic. In other words, in both the equilibrium and nonequilibrium
statistical mechanics the real relevant variables of interest are the properly \emph{averaged} (or time-smoothed) set 
of variables. 
%
%
%
\section{The Method of  Nonequilibrium Statistical Operator} 
%
%
The method of nonequilibrium statistical operator (NSO) developed by  Zubarev~\cite{zub74}
is the  satisfactory and workable approach to  construction of  Gibbs-type ensembles for  
nonequilibrium systems. The NSO method permits one to generalize the Gibbs ensemble method  to the nonequilibrium case naturally, and to
construct a nonequilibrium statistical operator which  enables one to obtain the transport 
equations and calculate the kinetic coefficients in terms of correlation functions, and which, 
in the case of equilibrium, goes over to the Gibbs distribution.\\ 
The NSO method sets out  as follows. The irreversible processes which can be considered as a reaction
of a system on mechanical perturbations can be analyzed by means of the method of linear reaction
on the external perturbation~\cite{zub74,kuzem17,kuzem07}. However, there is also a class of irreversible
processes induced by  thermal perturbations due to the internal inhomogeneity of a system. Among them
we have, e.g., diffusion, thermal conductivity, and viscosity. In   certain approximate schemes it is
possible to express such processes by  mechanical perturbations which artificially induce similar 
nonequilibrium processes. However, the fact is that the division of perturbations into mechanical and thermal ones is
reasonable in the linear approximation only~\cite{kuzem17}. In the higher approximations in the perturbation, 
mechanical perturbations can lead effectively to the appearance of  thermal perturbations.\\ The
NSO method permits one to formulate a workable scheme for description of the statistical mechanics of
irreversible processes which includes the thermal perturbation in a unified and coherent fashion.
To perform this, it is necessary to construct  statistical ensembles representing the macroscopic conditions
determining the system. 
Such a formulation is quite reasonable if we consider our system for a suitable large time.
For these large times the particular properties of the initial state of the system are irrelevant and
the relevant number of variables necessary for  description of the system reduces substantially.\\
A central issue in the statistical thermodynamics is the quest for the state functions that describe the changes of all
relevant (measurable) equilibrium quantities in terms of a set suitable state variables (thermodynamic state variables), that is, a set
of variables that uniquely determine a thermodynamic state. Equilibrium thermodynamics is based on two laws, each of which identifies
such a state functions. For the nonequilibrium thermodynamics the problem of a suitable choice of the relevant variables is much more
complicated.\\
The assumption of \emph{local equilibrium} is a basic and necessary assumption in linear irreversible 
thermodynamics~\cite{zub74,kuzem17,kuzem07,kuzrnc18,demi14}. It enables us to apply the equations of
equilibrium thermodynamics, such as the Gibbs equation, to local volume elements in a system. The entropy and other thermodynamic properties of
the system can then be defined in terms of local, intensive state variables. The assumption leads 
to the concept of an \emph{entropy production} in a system
subject to irreversible processes~\cite{zub74,kuzrnc18,msuz11,msuz11a,msuz12,espo10,mart06}.\\ 
Validity conditions for partial and complete local thermodynamic equilibrium in many-particle system play an important role
in the field of equilibrium and nonequlibrium statistical mechanics. 
A physical system is in an \emph{equilibrium state} if all currents - of heat, momentum, etc., - vanish, and the system is uniquely
described by a set of state variables, which do not change with time.\\
From a general point of view, all laws and thermodynamic relations for complete thermodynamic equilibrium also hold in the case of complete
local thermodynamic equilibrium. However, the only exception from this rule makes Planck 
radiation law~\cite{galga72,lgalga00,galg01,carati06}. 
In connection with the foundations of statistical mechanics,
the relations between thermodynamics and dynamics was considered in the context of the fact
that in quantum mechanics, equipartition should be replaced by Planck's  
law~\cite{galga72,lgalga00,galg01,carati06}. \\ But even in those cases in which
complete local thermodynamic equilibrium  does not further exist, partial local thermodynamic equilibrium (quasi-equilibrium) may still be realized, thus, permitting
nevertheless useful applications of general thermodynamic formulas under restricted conditions.\\
A case of considerable practical interest in connection with the phenomena of nonequilibrium processes is that of
the \emph{hierarchy of time scales}. One of the
essential virtues of the NSO method is that it focuses attention, at the outset, on the existence of different time scales.
Suppose that the Hamiltonian of our system can be divided as $ H = H_{0} + V$, where $ H_{0}$ is the dominant
part, and $V$ is a weak perturbation. The separation of the Hamiltonian into $ H_{0}$ and $V$ is not unique and 
depends on the physical properties of the system under consideration. The choice of the operator $ H_{0}$ determines
a short time scale $\tau_{0}$. This choice is such that for times $t \gg \tau_{0}$ the nonequilibrium state of the 
system can be described with a reasonable accuracy by the average values of some finite set of the operators 
$P_{m}$.\\
After the short time $\tau_{0}$, it is supposed that  the system can  achieve the state of an incomplete or 
quasi-equilibrium state.
The main assumption about the  quasi-equilibrium state is that it is determined completely by the quasi-integrals
of motion which are the internal parameters of the system. 
The characteristic relaxation time of these
internal parameters is much longer than $\tau_{0}$. Clearly then, that even if  these quasi-integrals  at 
the initial moment had no  definitive equilibrium values, after the time $\tau_{0}$, at the quasi-equilibrium state,
those parameters which  altered quickly became the functions of the external parameters and of the \emph{quasi-integrals of  motion}. 
It is essential that this functional connection does  not depend on the initial values of the parameters. In other words,
the operators $P_{m}$ are chosen so that they should satisfy the condition 
\begin{equation}\label{e4.1}
 [P_{k} , H_{0} ] = \sum_{l} c_{kl}P_{l}.
\end{equation}
It is necessary to write down the transport equations  for this set of \emph{relevant} operators only. 
The relevant operators may be scalars or vectors. The equations
of motion for the average of other   \emph{irrelevant}  operators (other physical variables) will be in some sense   
consequences of these transport equations.
As for the  \emph{irrelevant} operators which do not belong to the reduced set of the \emph{relevant} operators $P_{m}$, 
relation (\ref{e4.1}) leads to the infinite chain of operator equalities.  For times
$t \leq \tau_{0}$ the nonequilibrium averages of these operators oscillate fast, while for  times $t > \tau_{0}$ they become 
functions of the average values of the operators.\\
To carry out into practice the statistical thermodynamics of irreversible processes so that thermal perturbations were included, it is
necessary to construct a statistical ensemble representing the macroscopic conditions for the system~\cite{zub74}. 
For the construction of  a nonequilibrium statistical operator~\cite{zub74}
the basic hypothesis is that after small
time-interval $\tau$ the nonequilibrium distribution is established. Moreover, it is supposed that
it is weakly time-dependent by means of its parameter only. Then the statistical operator $\rho$ for
$t \geq \tau$ can be considered as an \emph{integral of motion} of the quantum Liouville equation
\begin{equation}\label{e4.2}
 \frac{\partial \rho }{\partial t} + \frac{1}{i \hbar}[\rho, H ] = 0.
\end{equation}
Here $\partial \rho / \partial t $ denotes time differentiation with respect to the time variable
on which the relevant parameters $F_{m}$ depend. It is important to note once again that $\rho$ depends
on $t$ by means of $F_{m}(t)$ only. These parameters are given through the external conditions for our system  and,
therefore, the term $\partial \rho / \partial t $ is the result of the external influence upon the system; this influence
causes that the system is non-stationary. In other words we may consider that the system is in thermal, material, and mechanical contact
with a combination of thermal baths and reservoirs  maintaining the given distribution of  parameters
$F_{m}$.  For example,  it can be the densities of energy, momentum, and particle  number for the
system which is macroscopically defined by given fields of temperature, chemical potential and
velocity. It is assumed that the chosen set of parameters is sufficient to characterize  macroscopically
the state of the system. Thus the choice of the set of the relevant parameters are dictated by the external
conditions for the system under consideration.\\ 
In order to describe the nonequilibrium process, it is supposed that the reduced set of variables incoming into $\rho $ is
chosen as the average value of some reduced set of relevant operators $P_{m}$, where $m$ is the index (continuous 
or discrete). For the suitable choice of operators $P_{m}$ such approach is possible for hydrodynamic and kinetic stage of
the irreversible process.\\
The equations of motions for  $P_{m}$ will lead to the suitable 
\emph{evolution equations}~\cite{zub74}. In the quantum case we have:
\begin{equation}\label{e4.3}
\frac{\partial P_{m}(t)}{\partial t} - \frac{1}{i \hbar}[P_{m}(t) , H ] = 0. 
\end{equation}
The time argument of the operators $P_{m}(t)$ denotes the Heisenberg
representation with the Hamiltonian $H$ independent of time. Then we suppose that the state of the ensemble
is  described by a nonequilibrium statistical operator which is a functional of $P_{m}(t)$
\begin{equation}\label{e4.4}
  \rho(t ) = \rho \{\ldots P_{m}(t)  \ldots \}.
\end{equation}
For the description of the hydrodynamic  stage of the irreversible process the energy, momentum and number of particles densities,
$H(x)$, $\mathbf{p}(x)$, $n_{i}(x)$ should be chosen as the operators $P_{m}(t)$. For the description of the kinetic  stage the occupation
number of one-particle states can be chosen.
It is necessary to take into account that $\rho(t)$ satisfies the Liouville equation.\\ 
Hence the  quasi-equilibrium (local-equilibrium)
Gibbs-type distribution will have the form
\begin{equation}\label{e4.5}
 \rho_{q} = Q^{-1}_{q} \exp \left(  - \sum_{m}F_{m}(t)P_{m}\right),
\end{equation}
where the parameters $F_{m}(t)$ have the sense of time-dependent thermodynamic parameters,
e.g., of temperature, chemical potential, and velocity (for the hydrodynamic stage), or the
occupation numbers of one-particle states (for the kinetic stage). The statistical
functional $Q_{q}$ is defined by demanding that the operator $\rho_{q}$ be normalized and  equal to
\begin{equation}\label{e4.6}
 Q_{q} = \textrm{Tr} \exp \left(  - \sum_{m}F_{m}(t)P_{m}\right).
\end{equation}
In addition, it was shown  that there exists general method for choosing a suitable
quasi-equilibrium distribution~\cite{zub74}. For the state with the extremal
value of the \emph{informational entropy}~\cite{zub74,kuzem17,jayn03,koz06,kuzem16,mart06}
\begin{equation}\label{e4.7}
  S = - \textrm{Tr} (\rho \ln \rho),
\end{equation}
provided that
\begin{equation}\label{e4.8}
   \textrm{Tr} ( \rho P_{m} ) = \langle P_{m}\rangle_{q}; \quad  \textrm{Tr}   \rho  = 1,
\end{equation}
it is possible to construct a suitable quasi-equilibrium ensemble~\cite{zub69,zub70,zuk70}.
Here the notation used is  $ \langle \ldots \rangle_{q} =  \textrm{Tr} ( \rho_{q}  \ldots )$. 
Then the corresponding quasi-equilibrium (or local equilibrium) distribution  has
the form~\cite{zub69,zub70,zuk70}
\begin{align}\label{e4.9}
\rho_{q}   =   \exp \left( \Omega - \sum_{m}F_{m}(t)P_{m}\right) \equiv \exp (- S(t,0)),\quad
\Omega = \ln \textrm{Tr}  \exp \left( - \sum_{m}F_{m}(t)P_{m}\right),
\end{align}
where $S(t,0)$ can be called  the entropy operator.
Indeed, the conditional extremum~\cite{zub69,zub70,zuk70} of the functional (\ref{e4.7}) corresponds to the extremum of
\begin{equation}\label{e4.10}
  \Phi (\rho) = - \textrm{Tr} ( \rho \ln \rho) - \sum_{m}F_{m} \textrm{Tr}(\rho P_{m}) +    \lambda \textrm{Tr}\rho,
\end{equation}
where $F_{m}(t)$ and $\lambda$ denote Lagrange multipliers. From the condition
\begin{equation}\label{e4.11}
 \delta \Phi ( \rho ) = 0,
\end{equation}
we find the expression for $\rho_{q}.$\\ 
The quasi-equilibrium statistical operator preserves the thermodynamic formulae for the parameters $F_{m}(t)$
\begin{equation}\label{e4.12}
 \frac{\delta \Phi}{\delta F_{m}}  =  - \langle P_{m}\rangle_{q}, 
\end{equation}
but the Liouville equation is not satisfied.\\
In other words, the form of the quasi-equilibrium statistical operator was constructed in such a way that to
ensure  that the thermodynamic equalities for the relevant parameters $F_{m}(t)$
\begin{equation}\label{e4.13}
  \frac{\delta \ln Q_{q}}{\delta F_{m}(t)} = \frac{\delta \Omega}{\delta F_{m}(t)} =  - \langle P_{m}\rangle_{q};
\quad \frac{\delta S}{\delta \langle P_{m}\rangle_{q} }  =  F_{m}(t) 
\end{equation}
are satisfied. 
It is clear that the variables $ F_{m}(t)$ and $\langle P_{m}\rangle_{q} $ are thermodynamically
conjugate. 
Since that the operator $\rho_{q}$ itself  does not satisfy the
Liouville equation, it should be modified~\cite{zub74} in such a way that the
resulting statistical operator satisfies the Liouville equation. This is the most delicate and subtle point
of the whole method. To clarify this point let us modify the quasi-equilibrium operator such that the Liouville equation
would be satisfied with the accuracy up to $\varepsilon \rightarrow 0.$ If we shall simply look for the statistical operator, which in a 
some initial moment is equal to the quasi-equilibrium operator, then, if the initial moment is fixed, we will have the transition effects
for small time intervals. These effects has not any real physical meaning. This is why Zubarev~\cite{zub74} used  
another way, remembering the averaging method in the nonlinear mechanics, which has much in common with the statistical 
mechanics. As it was pointed above,   
if the nonlinear system  tends to the limiting cycle it "\emph{forget}" about the initial conditions, as well as in the statistical mechanics. Thus,
according to Zubarev, the suitable variables (\emph{relevant operators}), which are time-dependent by means of $F_{m}(t)$, 
should be constructed by means of taking the \emph{invariant part} of the operators incoming into the logarithm of the statistical operator with respect to
the motion with Hamiltonian $H.$
Thus, by definition a special set of operators should be constructed which depends
on the time through the parameters $F_{m}(t)$ by taking the \emph{invariant part} of the operators
$F_{m}(t)P_{m}$ occurring in the logarithm of the quasi-equilibrium distribution, i.e.,
\begin{eqnarray}\label{e4.14}
 B_{m}(t) = \overline{F_{m}(t)P_{m}} = \varepsilon \int^{0}_{-\infty} e^{\varepsilon t_{1}}
F_{m}(t+ t_{1})P_{m}(t_{1})dt_{1} = \\
F_{m}(t)P_{m} - \int^{0}_{-\infty} dt_{1} e^{\varepsilon t_{1}}\left(F_{m}(t+ t_{1})\dot{P}_{m}(t_{1}) +
\dot{ F}_{m}(t+ t_{1})P_{m}(t_{1}) \right),\nonumber
\end{eqnarray}
where $(\varepsilon \rightarrow 0)$ and 
$$ \dot{P}_{m} =  \frac{1}{i \hbar}[P_{m}, H ]; \quad  \dot{ F}_{m}(t) = \frac{d F_{m}(t)}{dt}.$$
The parameter $\varepsilon > 0$ will be set
equal to zero, but only \emph{after the thermodynamic limit}~\cite{kuzem14} has been taken. Thus, the invariant part is taken
with respect to the motion with Hamiltonian $H$. 
The operators $B_{m}(t)$ satisfy the Liouville equation  in the limit $(\varepsilon \rightarrow 0)$
\begin{equation}\label{e4.15}
\frac{\partial B_{m}  }{\partial t} - \frac{1}{i \hbar}[ B_{m}, H ] =  
\varepsilon \int^{0}_{-\infty}  dt_{1}e^{\varepsilon t_{1}}\left(F_{m}(t+ t_{1})\dot{P}_{m}(t_{1}) +
\dot{ F}_{m}(t+ t_{1})P_{m}(t_{1}) \right).  
\end{equation}
The operation of taking the invariant part, or \emph{smoothing} the oscillating terms, is used in the
formal theory of scattering to set the boundary conditions which exclude the advanced solutions
of the Schr\"{o}dinger equation~\cite{zub74,kuzem17}. It is most clearly seen when the parameters $F_{m}(t)$ are
independent of time.   Differentiating
$\overline{P_{m}}$ with respect to time gives
\begin{equation}\label{e4.16}
\frac{\partial \overline{ P_{m}(t)}  }{\partial t}  =    \varepsilon \int^{0}_{-\infty} e^{\varepsilon t_{1}}
\dot{P}_{m}(t+t_{1})dt_{1}.  
\end{equation}
The $\overline{P_{m}(t)}$ can be called the integrals (or \emph{quasi-integrals}) of motion, although
they are conserved only in the limit $(\varepsilon \rightarrow 0)$. It is clear that for the Schr\"{o}dinger equation
such a procedure excludes the advanced solutions by  choosing the initial conditions. In the
present context this procedure leads to the selection of the retarded solutions of the Liouville 
equation.\\
The choice of the exponent in the statistical operator can be confirmed by considering its 
extremum properties~\cite{zub74,zub69,zub70,zuk70}. The requirement is that
the statistical operator should satisfy the condition of the minimum of the information 
entropy provided that
\begin{equation}\label{e4.17}
 \langle P_{m}(t_{1}) \rangle^{t+ t_{1}} =  \textrm{Tr} ( \rho_{q}  P_{m}(t_{1}) )  ); \quad   \textrm{Tr}  \rho_{q} = 1,
\end{equation}
in the interval $(- \infty \leq t_{1} \leq 0),$ i.e. for all moments of the past and with the preserved normalization. To this conditional extremum corresponds the
extremum of the functional~\cite{zub74,zub69,zub70,zuk70}
\begin{equation}\label{e4.18}
  \Phi ( \rho ) = - \textrm{Tr} ( \rho \ln \rho) -  \int^{0}_{-\infty} dt_{1}  \sum_{m}G_{m}(t_{1}) \textrm{Tr}(\rho P_{m}(t_{1})) +    \lambda \textrm{Tr}\rho,
\end{equation}
where $G_{m}(t_{1})$ and $\lambda$ are Lagrange multipliers. From the extremum condition it follows that
\begin{equation}\label{e4.19}
 \delta \Phi ( \rho ) =  - \textrm{Tr} (\delta \rho \ln \rho)  - \textrm{Tr} (\delta \rho ) +  \lambda \textrm{Tr} (\delta \rho ) -
\int^{0}_{-\infty} dt_{1}  \sum_{m}G_{m}(t_{1}) \textrm{Tr}(\delta \rho P_{m}(t_{1})) = 0,
\end{equation}
hence
\begin{equation}\label{e4.20}
 \rho =   \exp \bigl (  \Lambda - \int^{0}_{-\infty} dt_{1}  \sum_{m}G_{m}(t_{1}) P_{m}(t_{1})  \bigr); \quad   \Lambda = 1 - \lambda.
\end{equation}
Lagrange multipliers are determined by the conditions (\ref{e4.17}). 
We have
\begin{equation}\label{e4.21}
   \frac{\delta \tilde{\lambda}}{\delta G_{m}(t_{1})} =  \langle P_{m}(t_{1}) \rangle^{t}   = - \langle P_{m} \rangle^{t+ t_{1}}.
\end{equation}
If $P_{m}$ are integrals of motion, then the statistical operator $\rho$ (\ref{e4.20}) should give the Gibbs distribution, i.e. integral $\int^{0}_{-\infty} dt_{1}  \sum_{m}G_{m}(t_{1})$
should be convergent to a constant. It can be obtained if we put $G_{m}(t_{1}) = \varepsilon  e^{\varepsilon t_{1}}F_{m}. $ Taking into account
this property and the relation (\ref{e4.14}) we get that it is convenient to choose Lagrange multipliers in the form~\cite{zub74}
\begin{equation}\label{e4.22}
 G_{m}(t_{1}) = \varepsilon  e^{\varepsilon t_{1}}F_{m}(t + t_{1}).
\end{equation}
Then we shall obtain the statistical operator in the form (\ref{e4.20}), which corresponds to the extremum of the information entropy for a given
average $\langle P_{m} \rangle^{t_{1}}$ in an arbitrary moment of the past. 
The above consideration shows that the nonequilibrium statistical operator $\rho$ can be written as
\begin{eqnarray}\label{e4.23}
 \rho =  \exp (\overline{\ln \rho_{q}}) = \exp \Bigl (  \varepsilon \int^{0}_{-\infty}  dt_{1}e^{\varepsilon t_{1}} 
\exp \Bigl(\frac{iHt_{1}}{\hbar} \Bigr) \ln \rho_{q}(t+t_{1}) \exp \Bigl( \frac{- iHt_{1}}{\hbar} \Bigr) \Bigr ) = \\ \nonumber
\exp \Bigl (- \overline{S(t,0)} \Bigr ) = \exp \Bigl(- \varepsilon \int^{0}_{-\infty}  dt_{1}e^{\varepsilon t_{1}}S(t + t_{1},t_{1}) \Bigr)\\ \nonumber =   
\exp \Bigl ( - S(t,0) +  \int^{0}_{-\infty}  dt_{1}e^{\varepsilon t_{1}} \dot{S}(t + t_{1},t_{1}\bigr ) \Bigr ).
\end{eqnarray}
Here
\begin{eqnarray}\label{e4.24}
 \dot{S}(t,0) = \frac{\partial S(t,0)  }{\partial t} + \frac{1}{i \hbar}[S(t,0) , H ]; \quad   
 \dot{S}(t,t_{1}) = \exp \Bigl(\frac{iHt_{1}}{\hbar}\Bigr ) \dot{S}(t,0) \exp \Bigl(\frac{- iHt_{1}}{\hbar} \Bigr).
\end{eqnarray}
It is required~\cite{zub74} that the normalization of statistical operator $\rho_{q}$ is preserved as well as the statistical 
operator $\rho,$ and the  constraint $ \langle P_{m} \rangle^{t} =  \langle P_{m} \rangle^{t}_{q}$ is fulfilled.
For the particular choice of  $ F_{m}$ which corresponds to the statistical equilibrium
 we obtain $\rho = \rho_{q} = \rho_{0}$ and $\Lambda = \lambda$. It determines the parameters $ F_{m}(t)$ such that $ P_{m}$ and
$ F_{m}(t)$ are thermodynamically conjugate, i.e.
\begin{equation}\label{e4.28}
 \frac{\delta \lambda}{\delta F_{m}}  =  - \langle P_{m}\rangle_{q} =  - \langle P_{m}\rangle.
\end{equation}
It should be noted that a close related consideration  can also be carried out with a  deeper
concept, the methods of quasiaverages~\cite{zub74,qbog,zubar70}.
Zubarev showed~\cite{zubar70}  that the concepts of symmetry
breaking perturbations and quasiaverages~\cite{qbog,kuzem10} play 
important role in the theory of irreversible processes as
well. The method of the construction of the nonequilibrium
statistical operator becomes especially
deep and transparent when it is applied in the framework
of the quasiaverage concept. The main idea of
this approach was to consider infinitesimally small sources breaking the time-reversal symmetry of
the Liouville equation, which become vanishingly small after a thermodynamic 
limiting transition.\\  
Let us emphasize once again that the quantum Liouville equation, like the classical one, is symmetric under
time-reversal transformation. However, the solution of the Liouville equation is unstable with respect 
to small perturbations violating this symmetry of the equation. Indeed, let us consider the Liouville equation
with an infinitesimally small source into the right-hand side
\begin{equation}\label{e4.29}
 \frac{\partial \rho_{\varepsilon}  }{\partial t} + \frac{1}{i \hbar}[\rho_{\varepsilon}, H ] = 
 - \varepsilon ( \rho_{\varepsilon}  -  \rho_{q}),
\end{equation}
or equivalently
\begin{equation}\label{e4.30}
 \frac{\partial \ln \rho_{\varepsilon}  }{\partial t} + \frac{1}{i \hbar}[ \ln \rho_{\varepsilon}, H ] = 
 - \varepsilon ( \ln \rho_{\varepsilon}  - \ln \rho_{q}),
\end{equation}
where $(\varepsilon \rightarrow 0)$ after the thermodynamic limit. The equation (\ref{e4.29}) is analogous
to the corresponding equation of the quantum scattering theory~\cite{zub74,zubar70}. The introduction
of infinitesimally small sources into  the Liouville equation is equivalent to the boundary condition
\begin{equation}\label{e4.31}
  \exp \Bigl(\frac{iHt_{1}}{\hbar}\Bigr )\left( \rho(t + t_{1})  -  
  \rho_{q}(t + t_{1})  \right) \exp \Bigl(\frac{-iHt_{1}}{\hbar}\Bigr ) \rightarrow 0,
\end{equation}
where $t_{1} \rightarrow -\infty$ \emph{after} the thermodynamic limiting process. It was shown~\cite{zub74,zubar70}
that  we can rewrite the nonequilibrium statistical operator in the following useful form:
\begin{equation}\label{eq18d}
    \rho (t,0) =  \exp \left(
 - \varepsilon \int^{0}_{-\infty} dt_{1} e^{\varepsilon t_{1}} \ln \rho_{q}(t + t_{1},  t_{1})  \right) = 
 \exp \overline{\left( \ln \rho_{q}(t,0)\right)} \equiv \exp \overline{\left( -S(t,0)\right)}.
\end{equation}
The average value of any dynamic variable $A$ is given by
\begin{equation}\label{eq18e}
  \langle A \rangle = \lim_{\varepsilon \rightarrow 0^{+}} \textrm{Tr} ( \rho (t,0) A ),
\end{equation}
and is, in fact, the quasiaverage. The normalization of the quasi-equilibrium distribution $\rho_{q}$ will
persists after taking the invariant part if the following conditions will be fulfilled
\begin{equation}\label{eq18f}
  \textrm{Tr} ( \rho (t,0) P_{m} )= \langle P_{m}\rangle = \langle P_{m}\rangle_{q} ; \quad  \textrm{Tr}   \rho  = 1.  
\end{equation}
A short remark about the maximum entropy principle~\cite{zub74,kuzem17,jayn03,koz06,kuzem16,mart06} will not be out
of the place here. The approach to the nonequilibrium statistical mechanics which is based on the nonequilibrium ensembles
is related deeply with that principle, which was used in the  NSO method  as  one of the foundational issue. There are a few
slightly different equivalent possibilities of using the maximum entropy principle in this context, which were discussed thoroughly 
in Refs.~\cite{aus84,aus03}.
%
%
\section{Generalized Kinetic Equations}
%
%
It is well known that  kinetic equations are of great interest in the theory of transport
processes~\cite{bogol46,bog62,zub74,kuzem17,kuzem07,kuzem18}. The dynamic  behavior of charge,  magnetic, and lattice systems 
is of interest for the study of transport processes in solids~\cite{kuzem17,kuzem07,kuzem05,kuzem11,kuzem18}.
The degrees of freedom in solids   can  often be represented  as  a few interacting subsystems (electrons, 
spins, phonons, nuclear spins, etc.).  Perturbation of one subsystem may produce a nonequilibrium state which 
is then  relaxed to an equilibrium state due to the interaction between  particles or with a thermal bath.\\
The method of the nonequilibrium statistical operator  is a very useful tool
to analyze and derive   generalized transport and kinetic equations~\cite{zub74,kuzem07,kuzem05,kuzem11,pokrov2}.
The generalized kinetic equations  were derived by this method by  Pokrovski~\cite{zub74,pokrov2} 
for the case of many-particle system with small interactions among  particles. 
Indeed, as it was shown in the preceding section, the main quantities involved are
the following thermodynamically conjugate values:
\begin{equation}\label{e5.1}
 \langle P_{m}\rangle = - \frac{\delta \Omega}{\delta F_{m}(t)};
\quad F_{m}(t)  = \frac{\delta S}{\delta \langle P_{m}\rangle }.   
\end{equation}
The generalized transport equations which describe the \emph{time evolution} of variables $\langle P_{m}\rangle$ and $F_{m}$
follow from the equation of motion for the $ P_{m}$, averaged with the nonequilibrium statistical
operator (\ref{eq18d}). These equations have the form
\begin{equation}\label{e5.2}
 \langle \dot {P}_{m}\rangle = - \sum_{n} \frac{\delta^{2} \Omega}{\delta F_{m}(t)\delta F_{n}(t)}\dot{F}_{n}(t);
\quad \dot{F}_{m}(t)  =  \sum_{n} \frac{\delta^{2} S}{\delta \langle P_{m}\rangle \delta \langle P_{n}\rangle} \langle\dot {P}_{n}\rangle.  
\end{equation}
The entropy production has the form
\begin{equation}\label{e5.3}
\dot {S}(t) = \langle \dot {S}(t,0)\rangle = - \sum_{m}\langle \dot {P}_{m}\rangle F_{m}(t) = -
\sum_{n,m}\frac{\delta^{2} \Omega }{\delta F_{m}(t)\delta F_{n}(t)}\dot{F}_{n}(t)F_{m}(t).
\end{equation}
These equations are the mutually conjugate and  with   Eq.(\ref{eq18d}) form a complete system of equations
for the calculation of values  $\langle P_{m}\rangle$ and $F_{m}$.\\
Within the NSO method the derivation of the kinetic equations  for a system of 
weakly interacting particles was carried out by   Pokrovski~\cite{pokrov2}. In this case the Hamiltonian 
can be written in the form
\begin{equation}\label{e5.4}
 H = H_{0} + V,
\end{equation}
where $H_{0}$ is the Hamiltonian of noninteracting particles (or quasiparticles) and $V$ is the
operator describing the weak interaction among them. Let us choose the set of operators $ P_{m} = P_{k}$
whose average values correspond to the particle distribution functions, e.g.,  $a^{\dagger}_{k}a_{k}$ or
$a^{\dagger}_{k}a_{k+q}$. Here $a^{\dagger}_{k}$ and $a_{k}$ are the creation and annihilation second
quantized operators (Bose or Fermi type). These operators obey the following quantum equation of
motion: $\dot{P}_{k} = 1/i \hbar [P_{k} , H ].$
The averaging of this equation with NSO  gives the
generalized kinetic equations for $\langle P_{k}\rangle$
\begin{equation}\label{e5.6}
 \frac{d \langle P_{k}\rangle  }{d t} = \frac{1}{i \hbar}\langle [  P_{k} , H ]\rangle =
 \frac{1}{i \hbar} \sum_{l} c_{kl}\langle P_{l}\rangle +  \frac{1}{i \hbar}\langle [  P_{k} , V ]\rangle.
\end{equation}
Hence the calculation of the r.h.s. of (\ref{e5.6})
leads to the explicit expressions for the "collision integral" (collision terms). 
Since the interaction is small, it is possible to rewrite Eq.(\ref{e5.6}) in the following form:
\begin{equation}\label{e5.7}
 \frac{d \langle P_{k}\rangle  }{d t} = L^{0}_{k} + L^{1}_{k} + L^{21}_{k} + L^{22}_{k},
 \end{equation}
where
\begin{equation}\label{e5.8}
 L^{0}_{k} =  \frac{1}{i \hbar} \sum_{l} c_{kl}\langle P_{l}\rangle _{q}; \qquad
%
%
 L^{1}_{k} =  \frac{1}{i \hbar} \langle [  P_{k} , V ]\rangle_{q},
\end{equation}
\begin{equation}\label{e5.10}
 L^{21}_{k} =  \frac{1}{ \hbar^{2}}  \int^{0}_{-\infty} dt_{1} e^{\varepsilon t_{1}} \langle [ V(t_{1}), [ P_{k} , V ]]\rangle_{q},
\end{equation}
\begin{equation}\label{e5.11}
 L^{22}_{k} =  \frac{1}{ \hbar^{2}}  \int^{0}_{-\infty} dt_{1} 
 e^{\varepsilon t_{1}} \langle [ V(t_{1}),  i \hbar \sum_{l} P_{l}   
 \frac{\partial   L^{1}_{k}(\ldots \langle P_{l}\rangle \ldots)  }{\partial  \langle P_{l}\rangle } ]\rangle_{q}.
\end{equation}
The higher order terms  proportional to the $V^{3}$, $V^{4}$, etc., can be derived straightforwardly.
%
%
%
%
\section{System in Thermal Bath} 
%
In papers~\cite{kuzem07,kuzem05,kuzem18,wkuz70}    the generalized kinetic equations for  the system weakly coupled to a 
thermal bath have been derived. Examples of such system can be an atomic (or molecular) system interacting with the electromagnetic
field it generates as with a thermal bath, a system of nuclear or electronic spins interacting with the lattice, etc.
The aim was to describe  the relaxation processes in two weakly interacting subsystems, one of which is in the
nonequilibrium state and the other is considered as a thermal bath. The concept of thermal bath or heat reservoir,
i.e.,  a system that has effectively an infinite number of degrees of freedom, was not formulated precisely. A 
standard definition of the thermal bath is a heat reservoir defining a temperature of the system environment.
From a mathematical point of view~\cite{nnb78,koz08}, a heat bath is something that gives a stochastic influence on the system
under consideration.  The problem of
a small system weakly interacting with a heat reservoir has various aspects. 
Basic to the derivation of a
transport equation for a small system weakly interacting with a heat bath is a proper introduction of  model assumptions. 
We are interested here in the problem of
derivation of the kinetic equations for a certain set of  average values 
(occupation numbers, spins, etc.)  which characterize the nonequilibrium state of the system.\\
The Hamiltonian of the total system was taken in the following form:
\begin{equation}\label{e6.1}
 H = H_{1} + H_{2} + V,
\end{equation}
where
\begin{equation}\label{e6.2}
 H_{1} = \sum_{\alpha} E_{\alpha}a^{\dagger}_{\alpha}a_{\alpha}; 
 \quad V = \sum_{\alpha,\beta}\Phi_{\alpha \beta}a^{\dagger}_{\alpha}a_{ \beta}, \quad \Phi_{\alpha \beta} = 
 \Phi^{\dagger}_{ \beta \alpha}.
\end{equation}
Here $H_{1}$ is the Hamiltonian of the small subsystem, and $a^{\dagger}_{\alpha}$ and $a_{\alpha}$ are the creation and annihilation second
quantized operators of quasiparticles in the small subsystem with energies $ E_{\alpha}$, $V$ is the
operator of the interaction between the small subsystem and the thermal bath, and $H_{2}$ is the
Hamiltonian of the thermal bath  which we do not write explicitly. The quantities $\Phi_{\alpha \beta}$ are
the operators acting on the thermal bath variables.\\ 
We assume that the state of this
system is determined completely by the set of averages
$\langle P_{\alpha \beta}\rangle = \langle a^{\dagger}_{\alpha}a_{ \beta}\rangle$ and the state of the thermal bath 
by $\langle H_{2}\rangle$,
 where $ \langle \ldots \rangle$ denotes the statistical average with the nonequilibrium statistical operator,
 which will be defined below.\\
We take the quasi-equilibrium statistical operator $\rho_{q}$ in
the form
\begin{eqnarray}\label{e6.3}
\rho_{q} (t) =  \exp (- S(t,0)), \quad  S(t,0) = \Omega(t) + 
\sum_{\alpha \beta }P_{\alpha \beta }F_{\alpha\beta }(t) + \beta H_{2}. 
\end{eqnarray}
Here $F_{\alpha\beta }(t)$ are the thermodynamic parameters conjugated with $P_{\alpha \beta }$, and $\beta$ is
the reciprocal temperature 
of the thermal bath; $\Omega = \ln \textrm{Tr} \exp (- \sum_{\alpha \beta }P_{\alpha \beta }F_{\alpha\beta }(t) - \beta H_{2}).$
The nonequilibrium statistical operator has the form 
\begin{eqnarray}\label{e6.4}
\rho (t) =  \exp (- \overline{S(t,0)}); \quad
\overline{S(t,0)} = \varepsilon \int^{0}_{-\infty} dt_{1} e^{\varepsilon t_{1}} 
\left( \Omega(t + t_{1}) +  \sum_{\alpha \beta }P_{\alpha \beta }F_{\alpha\beta }(t) + \beta H_{2}  \right).
\end{eqnarray}
The parameters $F_{\alpha\beta }(t)$ are determined from the condition 
$\langle P_{\alpha \beta }\rangle = \langle P_{\alpha \beta }\rangle_{q}$.\\ In the derivation of the kinetic equations we  use
the perturbation theory in a \emph{weakness of interaction}  and  assume that the equality 
$ \langle \Phi_{\alpha \beta}\rangle_{q} = 0$ holds, while  other terms can be added to the renormalized energy of
the subsystem. 
We now turn to the derivation of the kinetic equations. The starting point is the kinetic equations
in the following implicit form:
\begin{equation}\label{e6.5}
 \frac{d \langle P_{\alpha \beta }\rangle }{d t} =  \frac{1}{i \hbar}\langle [  P_{\alpha \beta  } , H ]\rangle =
 \frac{1}{i \hbar}(E_{\beta} - E_{\alpha})\langle P_{\alpha \beta  }\rangle +  \frac{1}{i \hbar}\langle [  P_{\alpha \beta} , V ]\rangle.
\end{equation}
We restrict ourselves to the second-order in powers of $V$ in calculating the r.h.s. of (\ref{e6.5}). 
Finally we obtain the kinetic equations for $\langle P_{\alpha \beta }\rangle$ in
the form~\cite{kuzem05,kuzem18,wkuz70}
\begin{equation}\label{e6.6}
 \frac{d \langle P_{\alpha \beta }\rangle }{d t} =  \frac{1}{i \hbar}(E_{\beta} - E_{\alpha})\langle P_{\alpha \beta  }\rangle -
 \frac{1}{\hbar^{2}} \int^{0}_{-\infty} dt_{1} e^{\varepsilon t_{1}} 
 \langle \left[[P_{\alpha \beta}, V], V(t_{1}) \right]\rangle_{q}.
\end{equation}
The last term of the right-hand side of Eq.(\ref{e6.6}) can be called the generalized  \emph{collision integral}.
Thus,  we can see that the collision term for the system  weakly coupled to the thermal bath has a convenient
form of the double commutator as for the generalized kinetic equations~\cite{pokrov2}  for the system with
small interaction. It should be emphasized  that the assumption about the model form of the Hamiltonian
(\ref{e6.1}) is nonessential for the above derivation~\cite{kuzem07,kuzem05,kuzem18,wkuz70}. 
The equation (\ref{e6.6}) will be fulfilled for  general form of the Hamiltonian of a small system weakly coupled to a
thermal bath.\\
Having shown the derivation of the generalized kinetic equations, we note in this context that
investigation of quantum dynamics in the condensed phase is one of a major objective of many recent studies~\cite{kuzem17,kuzem07,kuzem06}. One of the
useful approaches to quantum dynamics in the condensed phase is based on a reduced density matrix approach. An equation of motion
for the reduced density matrix was obtained by averaging out of the full density matrix irrelevant bath degrees of freedom which indirectly
appear in observations via coupling to the system variables. A well known standard reduced density matrix approach is 
the Redfield's equations, which were derived also within the NSO formalism~\cite{kuzem17,kuzem07,kuzem06}.
We  derived above the kinetic equations for $\langle P_{\alpha \beta }\rangle$ in
general form.  It is possible to rewrite the kinetic equations for $\langle P_{\alpha \beta }\rangle$ as
\begin{align}\label{e6.7}
 \frac{d \langle P_{\alpha \beta }\rangle }{d t} =  \frac{1}{i \hbar}(E_{\beta} - E_{\alpha})\langle P_{\alpha \beta  }\rangle - \nonumber \\
\sum_{\nu} \left( K_{\beta\nu}\langle P_{\alpha \nu }\rangle  +  K^{\dag}_{\alpha \nu} \langle P_{\nu \beta}\rangle \right) +
\sum_{\mu \nu} K_{\alpha \beta,\mu \nu} \langle P_{\mu \nu }\rangle.
\end{align}
For notation see Refs.~\cite{kuzem17,kuzem07}
The above result is similar in structure to the Redfield's equation for the spin density matrix~\cite{kuzem17} when
the external time-dependent field is absent. Indeed, the Redfield's equation of motion for the spin density matrix
has the form~\cite{kuzem17}
$$ \frac{  \partial \rho^{\alpha \alpha' }}{\partial t} = - i \omega_{\alpha \alpha'}\rho^{\alpha \alpha' } + 
\sum_{\beta \beta'} R_{\alpha \alpha' \beta \beta'}\rho^{\beta \beta' }.$$
Here $\rho^{\alpha \alpha' } $ is the $\alpha, \alpha'$ matrix element of the spin density matrix,
 $\omega_{\alpha \alpha'} =  (E_{\alpha} - E_{\alpha'}) \hbar  $, where $E_{\alpha}$ is energy of the spin state $\alpha$
and $R_{\alpha \alpha' \beta \beta'}\rho^{\beta \beta' } $ is the "relaxation matrix". A sophisticated analysis and
derivation of the Redfield's equation for the density of a spin system immersed in a thermal bath was given in~\cite{kuzem17,kuzem07}.\\ 
Returning to  Eq.(\ref{e6.7}), it is it can be shown, that if one confines himself to the diagonal averages $\langle P_{\alpha \alpha}\rangle$ only, 
this equation  may be transformed to give~\cite{kuzem17,kuzem07,kuzem18,wkuz70}
\begin{equation}\label{e6.8}
 \frac{d \langle P_{\alpha \alpha }\rangle }{d t} =  \sum_{\nu} K_{\alpha \alpha,\nu \nu} \langle P_{\nu \nu }\rangle
 - \left( K_{\alpha \alpha}  + K^{\dag}_{\alpha \alpha}  \right)\langle P_{\alpha \alpha }\rangle,
\end{equation}
\begin{eqnarray}\label{e6.9}
K_{\alpha \alpha,\beta \beta} =  \frac{1}{\hbar^{2}} J_{\alpha \beta, \beta \alpha} ( \frac{E_{\alpha} - E_{\beta}}{\hbar} ) = 
W_{\beta \rightarrow \alpha},\\
 K_{\alpha \alpha}  + K^{\dag}_{\alpha \alpha}  = 
 \frac{1}{\hbar^{2}}\sum_{\beta}  J_{\beta \alpha, \alpha \beta} ( \frac{E_{\beta} - E_{\alpha}}{\hbar} ) = 
W_{\alpha \rightarrow \beta}. 
\end{eqnarray}
Here $W_{\beta \rightarrow \alpha}$ and $W_{\alpha \rightarrow \beta}$ are the transition probabilities
expressed in the spectral intensity terms. Using the properties of the spectral intensities~\cite{zub74},
it is possible to verify that the transition probabilities satisfy the relation of the detailed
balance
\begin{equation}\label{e6.10}
\frac{W_{\beta \rightarrow \alpha}}{W_{\alpha \rightarrow \beta}} = 
\frac{\exp (-\beta E_{\alpha})}{\exp (-\beta E_{\beta})}.
\end{equation}
Finally, we have
\begin{equation}\label{e6.11}
 \frac{d \langle P_{\alpha \alpha }\rangle }{d t} =  \sum_{\nu} W_{\nu \rightarrow \alpha}  \langle P_{\nu \nu }\rangle
 - \sum_{\nu} W_{\alpha \rightarrow \nu}  \langle P_{\alpha \alpha}\rangle.
 \end{equation}
This equation has the usual form of the Pauli master equation.  It is  known that  the master equation
is an ordinary differential equation  describing the {\em reduced evolution} of the system  obtained from the full
Heisenberg evolution by taking the partial expectation with respect to the vacuum state of the reservoirs degrees
of freedom.
In this sense, the generalized master equation is a tool for extracting the dynamics of
a subsystem of a larger system  by the use of a special projection techniques~\cite{zwan01},  or special 
expansion technique~\cite{kuzem17,kuzem07}.
%
%
%
%
\section{Schr\"{o}dinger-Type Equation with Damping} 
%
%
The problem of the inclusion of dissipative forces in quantum mechanics
is of  great interest. There are various approaches to this complicated problem~\cite{kuzem17,kuzem07}.
The inclusion of dissipative forces in quantum mechanics
through the use of non-Hermitian Hamiltonians is of  great interest in the theory of  interaction between heavy ions.
It is clear that if the Hamiltonian has a non-Hermitian part $H_{A},$ the Heisenberg equation of motion will be modified
by  additional terms. However, care must be taken in defining the probability density operator when the 
Hamiltonian is non-Hermitian. The necessity of considering such  processes arises in the description of various 
quantum phenomena (e.g. radiation damping, etc.), since quantum systems experience dissipation and fluctuations 
through interaction with a reservoir~\cite{kuzem17,kuzem07,kuzem18}.
Here we consider the behavior of
a small dynamic  system interacting with a thermal bath, i.e., with a system that has effectively an infinite
number of degrees of freedom, in the approach of the nonequilibrium statistical operator~\cite{kuzem17,kuzem07,wzkuz70}, on the basis of the equations 
described in previous section. 
It was  assumed that the dynamic  system (system of particles) is far from
equilibrium with the thermal bath and cannot, in general, be characterized by a temperature. As a result of
interaction with the thermal bath, such a system   acquires some statistical characteristics but   remains
essentially a mechanical system. Our aim was to obtain an equation of evolution (equations of motion) for the
relevant variables which are characteristic of the system under consideration~\cite{kuzem17,kuzem07,wzkuz70}. The basic 
idea was to eliminate effectively the thermal bath variables.  The influence of the thermal bath
will be manifested then as an \emph{effect of friction} of the particle in a medium. The presence of friction leads to
dissipation and, thus, to irreversible processes. In other words, it was supposed that  
the reservoir can be {\em completely eliminated}, provided that the frequency shifts and dissipation
induced by the reservoir are incorporated into the properly averaged equations of motion, and provided that a suitable 
operator \emph{noise source} with the correct moments are added.\\
Let us consider the behavior of a small subsystem with Hamiltonian
$H_{1}$ interacting with a thermal bath with Hamiltonian $H_{2}$. The total Hamiltonian has the form 
(\ref{e6.1}). As operators $P_{m }$ determining the nonequilibrium state of the small subsystem, we take
$a^{\dagger}_{\alpha}, a_{\alpha}$, and $n_{\alpha} = a^{\dagger}_{\alpha}a_{\alpha}$. Note  that the choice
of only the operators $n_{\alpha}$ and $H_{2}$ would lead to kinetic equations (\ref{e6.6})  for the system in the thermal bath
derived above.\\
The quasi-equilibrium statistical operator (\ref{e6.3}) is determined from the extremum of the
information entropy, subjected to the additional conditions that the quantities
\begin{equation}\label{e7.1}
\textrm{Tr} (\rho a_{\alpha}) = \langle a_{\alpha}\rangle, \quad \textrm{Tr} (\rho a^{\dagger}_{\alpha}) = \langle a^{\dagger}_{\alpha}\rangle, \quad
\textrm{Tr} (\rho n_{\alpha}) = \langle n_{\alpha}\rangle
\end{equation}  
remain constant during the variation and the normalization $\textrm{Tr} (\rho ) = 1$ is preserved. The operator
$\rho_{q}$ has the form
\begin{eqnarray}\label{e7.2}
\rho_{q}   =   \exp \left( \Omega - \sum_{\alpha} ( f_{\alpha}(t)a_{\alpha} + f^{\dagger}_{\alpha}(t)a^{\dagger}_{\alpha} + 
F_{\alpha}(t)n_{\alpha} ) - \beta H_{2} \right)  \equiv \exp ( - S(t,0)). 
\end{eqnarray}
Here, $f_{\alpha}, f^{\dagger}_{\alpha}$ and $F_{\alpha}$ are Lagrangian multipliers determined by the conditions
(\ref{e7.1}). They are the parameters conjugate to $\langle a_{\alpha}\rangle_{q}, \langle a^{\dagger}_{\alpha}\rangle_{q}$  
and $\langle n_{\alpha}\rangle_{q}$:
\begin{equation}\label{e7.3}
\langle a_{\alpha}\rangle_{q} =  - \frac{\delta \Omega}{\delta f_{\alpha}(t)},  \quad \langle n_{\alpha}\rangle_{q} =
- \frac{\delta \Omega}{\delta F_{\alpha}(t)},
\quad \frac{\delta S}{\delta \langle a_{\alpha}\rangle_{q} }  =  f_{\alpha}(t),  \quad
\frac{\delta S}{\delta \langle n_{\alpha}\rangle_{q} } = F_{\alpha}(t). 
\end{equation}  
It is worth noting that our choice of the relevant operators $\langle a_{\alpha}\rangle_{q}, \langle a^{\dagger}_{\alpha}\rangle_{q}$ 
precisely corresponds to the ideas of the McLennan, described above. His method
is based on the introduction of external forces of a nonconservative nature, which describe the influence of the surroundings or of a thermal bath
on the given system. Indeed, our choice means introduction of artificial external forces, which broke the law of the particle conservation. This is
especially radical view for the case of Fermi-particles, since it broke the spin conservation law too~\cite{kuzpaw72}.\\
In what follows, it is convenient to write the quasi-equilibrium statistical operator (\ref{e6.3}) in the factorized
form $\rho_{q} = \rho_{1} \bigotimes \rho_{2}.$
Here the notation are:
\begin{eqnarray}\label{e7.4}
\rho_{1}   =   \exp \left( \Omega_{1} - \sum_{\alpha} ( f_{\alpha}(t)a_{\alpha} + f^{\dagger}_{\alpha}(t)a^{\dagger}_{\alpha} + 
F_{\alpha}(t)n_{\alpha} ) \right), \quad
\rho_{2}   =   \exp \left( \Omega_{2}  - \beta H_{2} \right). 
\end{eqnarray}
The nonequilibrium statistical operator $\rho$ will have the form (\ref{e6.4}). Note, that the following conditions
are satisfied:
\begin{equation}\label{e7.5}
\langle a_{\alpha}\rangle_{q} = \langle a_{\alpha}\rangle, \quad \langle a^{\dagger}_{\alpha}\rangle_{q} = \langle a^{\dagger}_{\alpha}\rangle, \quad
\langle n_{\alpha}\rangle_{q} = \langle n_{\alpha}\rangle.
\end{equation}  
We shall take, as our starting point, the equations of motion for the operators averaged with the
nonequilibrium statistical operator 
\begin{eqnarray}\label{e7.6}
i\hbar \frac{d \langle a_{\alpha}\rangle}{dt} = \langle [a_{\alpha}, H_{1} ]\rangle + \langle [a_{\alpha}, V ]\rangle, \quad
i\hbar \frac{d \langle n_{\alpha}\rangle}{dt} = \langle [n_{\alpha}, H_{1} ]\rangle + \langle [n_{\alpha}, V ]\rangle.
\end{eqnarray}
The equation for $\langle a^{\dagger}_{\alpha}\rangle$ can be obtained by taking the conjugate of  (\ref{e7.6}). Restricting
ourselves to the second order in the interaction V, we obtain, by analogy with (\ref{e6.6}), the following
equations~\cite{kuzem17,kuzem07,wzkuz70}:
\begin{eqnarray}\label{e7.8}
i\hbar \frac{d \langle a_{\alpha} \rangle}{dt} = E_{\alpha}\langle a_{\alpha}\rangle + \frac{1}{i\hbar }
\int^{0}_{-\infty} dt_{1} e^{\varepsilon t_{1}} 
 \langle \left[[a_{\alpha}, V], V(t_{1}) \right]\rangle_{q},\\
i\hbar \frac{d \langle n_{\alpha}\rangle}{dt} = 
\frac{1}{i\hbar }
\int^{0}_{-\infty} dt_{1} e^{\varepsilon t_{1}} 
 \langle \left[[n_{\alpha}, V], V(t_{1}) \right]\rangle_{q}.
\label{e7.9} 
\end{eqnarray}
Here $V(t_{1})$ denotes the interaction representation of the operator $V$. Expanding the double commutator
in Eq.(\ref{e7.8}), we obtain
\begin{align}\label{e7.10}
i\hbar \frac{d \langle a_{\alpha}\rangle}{dt} = E_{\alpha}\langle a_{\alpha}\rangle   \nonumber \\ +   \frac{1}{i\hbar }
\int^{0}_{-\infty} dt_{1} e^{\varepsilon t_{1}} \left( \sum_{\beta \mu \nu} 
\langle \Phi_{\alpha \beta}\phi_{\mu \nu}(t_{1})\rangle_{q}\langle a_{\beta}a^{\dagger}_{\mu} a_{\nu}\rangle_{q}   -     
\langle \phi_{\mu \nu}(t_{1})\Phi_{\alpha \beta}\rangle_{q}\langle a^{\dagger}_{\mu} a_{\nu} a_{\beta}\rangle_{q} \right),  
\end{align}
where $\phi_{\mu \nu}(t_{1}) = \Phi_{\mu \nu}(t_{1}) \exp \bigl ( i/\hbar(E_{\mu} - E_{\nu})t_{1} \bigr )$. 
We  assume that the terms of higher order than linear can be dropped. Then we get
\begin{equation}\label{e7.11}
  i\hbar \frac{d \langle a_{\alpha} \rangle}{dt} = E_{\alpha}\langle a_{\alpha}\rangle + \frac{1}{i\hbar }\sum_{\beta \mu }
\int^{0}_{-\infty} dt_{1} e^{\varepsilon t_{1}} \langle \Phi_{\alpha \mu}\phi_{\mu \beta}(t_{1})\rangle_{q}\langle a_{\beta}\rangle.
\end{equation}
The form of the linear equation (\ref{e7.11}) is the same for Bose and Fermi statistics.\\ Using the 
spectral representations~\cite{zub74,wzkuz70},  it is possible  to rewrite Eq.(\ref{e7.11}) 
by analogy with Eq.(\ref{e6.8}) as
\begin{equation}\label{e7.12}
  i\hbar \frac{d \langle a_{\alpha}\rangle}{dt} = E_{\alpha}\langle a_{\alpha}\rangle + \sum_{\beta }
K_{\alpha \beta} \langle a_{\beta}\rangle.
\end{equation}
%
Thus, we have obtained the equation of
motion for the average $\langle a_{\alpha}\rangle$. 
It is clear that this equation describes approximately the evolution of the state of the dynamic  system interacting with
the thermal bath. The last term in the right-hand side of this equation leads to the shift of 
energy $E_{\alpha}$ and to the damping due to the interaction  with the thermal bath (or medium). In a certain sense,
it is possible to say that  Eq.(\ref{e7.12}) is an analog or the generalization of the 
Schr\"{o}dinger equation~\cite{kuzem17,kuzem07,wzkuz70}.\\ 
It is of interest to analyze and   track more closely the analogy with the
Schr\"{o}dinger equation in the coordinate form. To do this, by convention, we define the "\emph{wave function}"
\begin{equation}\label{e7.13}
\psi ( \mathbf{r}) = \sum_{\alpha} \chi_{\alpha}( \mathbf{r})\langle a_{\alpha}\rangle,
\end{equation}
where $\{ \chi_{\alpha}( \mathbf{r}) \}$ is a complete orthonormalized system of single-particle 
functions of the operator
$\left( - \hbar^{2}/2m \nabla^{2}  + v(\mathbf{r}) \right)$,
where $v(\mathbf{r})$ is the potential energy, and
\begin{equation}\label{e7.14}
\left(- \frac{\hbar^{2}}{2m}\nabla^{2}  + v(\mathbf{r}) \right)\chi_{\alpha}( \mathbf{r}) = 
E_{\alpha} \chi_{\alpha}( \mathbf{r}).
\end{equation}
Thus, in a certain sense, the quantity $\psi ( \mathbf{r})$ may plays the role of the wave function of a  
particle in a medium. Now, using (\ref{e7.13}), we transform Eq.(\ref{e7.12}) to 
(see Refs.~\cite{kuzem07,kuzem05,kuzem18,wzkuz70})  
\begin{equation}\label{e7.15}
 i\hbar \frac{\partial \psi ( \mathbf{r}) }{\partial t} = 
 \left(- \frac{\hbar^{2}}{2m} \nabla^{2}  + v(\mathbf{r}) \right) \psi( \mathbf{r})  +
 \int K (\mathbf{r},\mathbf{r'})\psi ( \mathbf{r'})d \mathbf{r'}.
\end{equation}
The kernel $K (\mathbf{r},\mathbf{r'})$ of the integral equation (\ref{e7.15}) has the form
\begin{equation}\label{e7.16}
  K (\mathbf{r},\mathbf{r'}) = \sum_{\alpha \beta} K_{\alpha \beta}\chi_{\alpha}( \mathbf{r})\chi^{\dag}_{\beta}( \mathbf{r'}) = 
\frac{1}{i\hbar} \sum_{\alpha, \beta, \mu } \int^{0}_{-\infty} dt_{1} e^{\varepsilon t_{1}}
\langle \Phi_{\alpha \mu}\phi_{\mu \beta}(t_{1})\rangle_{q}\chi_{\alpha}( \mathbf{r})\chi^{\dag}_{\beta}( \mathbf{r'}).
\end{equation}
Equation (\ref{e7.15}) can be called a \emph{Schr\"{o}dinger-type equation} with damping for a dynamical system in
a thermal bath. It is interesting to note that similar Schr\"{o}dinger equations with a nonlocal interaction are used in the
scattering theory~\cite{kuzem17} to describe interaction with many scattering centers. \\To demonstrate the
capabilities of   equation (\ref{e7.15}), it is convenient to introduce the operator of translation
$ \exp (i\mathbf{q} \mathbf{p} / \hbar)$, where $\mathbf{q} = \mathbf{r'} - \mathbf{r}; \, \mathbf{p} = -i\hbar \nabla_{r}$.
Then Eq.(\ref{e7.15}) can be rewritten in the form
\begin{equation}\label{e7.17}
 i\hbar \frac{\partial \psi ( \mathbf{r}) }{\partial t} = 
 \left(- \frac{\hbar^{2}}{2m} \nabla^{2}  + v(\mathbf{r}) \right) \psi( \mathbf{r})  +
\sum_{p} D(\mathbf{r},\mathbf{p})\psi ( \mathbf{r}), 
\end{equation}
where
\begin{equation}\label{e7.18}
D(\mathbf{r},\mathbf{p}) = \int d^{3}q K (\mathbf{r},\mathbf{r} + \mathbf{q}) \exp \Bigl (  \frac{i\mathbf{q}\mathbf{p}}{\hbar} \Bigr ).
\end{equation}
It is reasonable to assume that the wave function $\psi ( \mathbf{r})$ varies little over the correlation 
length characteristic of the kernel $K (\mathbf{r},\mathbf{r'})$. Then, 
expanding $ \exp (i\mathbf{q} \mathbf{p} / \hbar)$ in a series, we obtain the following equation in the
zeroth order:
\begin{equation}\label{e7.19}
 i\hbar \frac{\partial \psi ( \mathbf{r}) }{\partial t} = 
 \left(- \frac{\hbar^{2}}{2m} \nabla^{2}  + v(\mathbf{r}) + \textrm{Re}   U( \mathbf{r}) \right) \psi( \mathbf{r})  +
 i \, \textrm{Im}   U( \mathbf{r}) \psi( \mathbf{r}),
\end{equation}
where
\begin{equation}\label{e7.20}
U( \mathbf{r}) =  \textrm{Re}  U( \mathbf{r}) +  i \, \textrm{Im}   U( \mathbf{r}) =
 \int d^{3}q K (\mathbf{r},\mathbf{r} + \mathbf{q}). 
\end{equation}
Expression (\ref{e7.19}) has the form of a Schr\"{o}dinger equation with a complex potential\index{Schrodinger equation with a complex potential}. 
Equations of this form are well known in the scattering theory~\cite{kuzem17},  in which one introduces an 
interaction describing absorption  $(\textrm{Im}  U( \mathbf{r})  < 0)$. 
Let us consider the expansion of $\exp \Bigl ( i\mathbf{q} \mathbf{p} / \hbar  \Bigr )$  in Eq.(\ref{e7.18})  in a series up to second order inclusively.
Then we can represent Eq.(\ref{e7.15}) in the form~\cite{kuzem07,kuzem18,wzkuz70} 
\begin{eqnarray}\label{e7.21}
i\hbar \frac{\partial \psi ( \mathbf{r}) }{\partial t} = 
\Bigl \{  \left(- \frac{\hbar^{2}}{2m} \nabla^{2}  + v(\mathbf{r}) \right)   +  U( \mathbf{r}) 
- \frac{1}{i\hbar} \int d \mathbf{r'}  K (\mathbf{r},\mathbf{r} + \mathbf{r'})\mathbf{r'} \mathbf{p} \\ \nonumber
+ \frac{1}{2} \int  K (\mathbf{r},  \mathbf{r} + \mathbf{r'})d \mathbf{r'}  \sum_{i,k =1}^{3} r'_{i} r'_{k} \nabla_{i} \nabla_{k} \Bigr \} \psi( \mathbf{r}).
\end{eqnarray}
Let us introduce  the function
\begin{equation}
\mathbf{A} (\mathbf{r}) = \frac{m c}{i\hbar e} \int  d \mathbf{r'} K (\mathbf{r},\mathbf{r} + \mathbf{r'}) \mathbf{r'},
\end{equation}
which, in a certain sense, is the analog of the complex vector potential of an electromagnetic field. Then we can define 
an analog of the tensor of the reciprocal effective masses, considered in details in 
Refs.~\cite{kuzem17,kuzem18,kuzem08}
\begin{equation}
\Bigl \{  \frac{1}{M (\mathbf{r})} \Bigr \}_{ik} = \frac{1}{m} \delta_{ik} - \int  d \mathbf{r'} \textrm{Re}  K (\mathbf{r},\mathbf{r} + \mathbf{r'}) r'_{i} r'_{k} .
\end{equation}
Hence   we can rewrite Eq.(\ref{e7.15}) in the form
\begin{eqnarray}\label{e7.22}
i\hbar \frac{\partial \psi ( \mathbf{r}) }{\partial t} \qquad  \\ \nonumber
= \Bigl \{ - \frac{\hbar^{2}}{2} \sum_{i,k} \Bigl ( \frac{1}{M (\mathbf{r})} \Bigr )_{ik} \nabla_{i} \nabla_{k} + v(\mathbf{r})    +  U( \mathbf{r})
+ \frac{i\hbar e}{m c} \mathbf{A} (\mathbf{r}) \nabla + i T (\mathbf{r}) \Bigr \} \psi( \mathbf{r}),
\end{eqnarray}
where
\begin{equation}
 T (\mathbf{r}) = \frac{1}{2} \int  d \mathbf{r'} \textrm{Im}  K (\mathbf{r},\mathbf{r} + \mathbf{r'})\sum_{i,k}  r'_{i} r'_{k} \nabla_{i} \nabla_{k}. 
\end{equation}
In the case of an isotropic medium, the tensor  $\Bigl \{ 1/M (\mathbf{r})\Bigr \}_{ik}$  is diagonal and $\mathbf{A} (\mathbf{r}) = 0.$\\
It is worth while to mention that the transmission and scattering problems involving complex
potentials are important in physics, in particular in describing nuclear collisions~\cite{kuzem17}.\\
Note, that the introduction of $\psi (\mathbf{r})$ does not mean that the state of the small dynamical subsystem is
pure. It remains mixed since it is described by the statistical operator (\ref{e7.4}), the evolution of the
parameters  $f_{\alpha}, f^{\dagger}_{\alpha}$, and $F_{\alpha}$  of the latter being governed by a 
coupled system of equations of Schr\"{o}dinger and kinetic types. \\
Hence, we have shown in this section  that for some class of dynamic 
systems it was possible,  with the NSO approach,  to go from a Hamiltonian description of dynamics to a description in terms
of processes which incorporates the dissipation~\cite{kuzem17,kuzem07,kuzem05,kuzem18,wzkuz70}. However, a careful examination is required in order to see under what conditions
the Schr\"{o}dinger-type equation with damping can really be used. 
%
%
%
\section{Damping Effects in Open Dynamical System}
%
%
In order to clarified this point and to interpret properly the physical meaning of the derived
equations, the example  will be considered here. Let us  consider briefly 
important example~\cite{lnc71} of the application of the Schr\"{o}dinger-type equation with damping.
We consider the problem of the natural width of spectral line of the atomic system and show that our result coincides with the results
obtained earlier by other methods. It is well known that the excited levels  of the isolated atomic system have a finite lifetime because 
there is a probability of emission of photons due to interaction with the self-electromagnetic field. This leads to the atomic levels
becoming quasi-discrete and consequently acquiring a finite small width. It is just this width that is called the \emph{natural width} 
of the spectral lines. \\
Let us consider an atom interacting with the self-electromagnetic field in the approximation when the atom is at rest. For simplicity, 
the atom is supposed to be in two states only, i.e. in a ground state $a$ and in an excited state $b.$ The atomic system in the excited state
$b$ is considered, in a certain sense, as a small "nonequilibrium" system, and the self-electromagnetic field as a "thermostat" 
or a "thermal bath".
The relaxation of the small system is then a decay of the excited level and occurs by radiative transitions\index{radiative transitions}.\\
We shall not discuss here the case when the electromagnetic field can be considered as an equilibrium system with infinitely many degrees
of freedom, because it has been discussed completely in the literature. 
We write the total Hamiltonian in the form $H = H_{at} + H_{f} + V,$ where
\begin{equation} 
 H_{at} = \sum_{\alpha} E_{\alpha}a^{\dagger}_{\alpha}a_{\alpha}
\end{equation}
is the Hamiltonian for the atomic system alone, $a^{\dagger}_{\alpha}$ and $a_{\alpha}$ are the creation 
and annihilation operators of the system in the state with energy $E_{\alpha}$.
\begin{equation} 
 H_{f} = \sum_{k, \lambda} k c b^{\dagger}_{k, \lambda}b_{k, \lambda}
\end{equation}
is the Hamiltonian of transverse electromagnetic field~\cite{kuzem17,lnc71}, $\lambda = 1,2$ is the 
polarization, $\hbar \mathbf{k}$ is the momentum of a photon, $b^{\dagger}_{k, \lambda}$ and $b_{k, \lambda}$ are the creation and annihilation operators of the photon in the state $(\mathbf{k} \lambda)$,
$c$ is the light velocity, $V$ is the interaction operator responsible for the radiative transitions and having the following form in the
non-relativistic approximation 
\begin{equation}
 V =    - \frac{e}{mc} \mathbf{p}\cdot \mathbf{A}_{tr}(\mathbf{r}),
\end{equation}
where $e$ and $m$ are the electron charge and mass, respectively, $\mathbf{A}_{tr}(\mathbf{r})$ is the vector-potential of the transverse electromagnetic field
at the point $\mathbf{r}$; $[ \mathbf{p} \times  \mathbf{A}_{tr}(\mathbf{r})] = 0.$ For a finite system enclosed in a cubic box of volume $\Omega$ with periodic
boundary conditions, one can write~\cite{kuzem17,lnc71}
\begin{eqnarray} 
\mathbf{A}_{tr}(\mathbf{r}) = \frac{1}{\sqrt{\Omega}}\sum_{k, \lambda} \left (  \frac{2 \pi \hbar^{2} c}{k} \right)^{1/2} \mathbf{e}_{k, \lambda}
\Bigl ( b_{k, \lambda} \exp (\frac{i \mathbf{k} \mathbf{r}}{\hbar} ) + b^{\dagger}_{k, \lambda} \exp ( - \frac{i \mathbf{k} \mathbf{r}}{\hbar} ) \Bigr ).
\end{eqnarray}
Now, following the derivation of section 7, the interaction $V$  is represented as a product, such that the atomic and 
field variables are factorized:
\begin{equation}
 \quad V = \sum_{\alpha,\beta}\varphi_{\alpha \beta}a^{\dagger}_{\alpha}a_{ \beta}, \quad \varphi_{\alpha \beta} = 
 \varphi^{\dagger}_{\beta \alpha },
\end{equation}
where
\begin{equation}
\varphi_{\alpha \beta} =  \frac{1}{\sqrt{\Omega}}\sum_{k, \lambda}
\Bigl ( G_{\alpha,\beta}(k, \lambda)  b_{k, \lambda}  + b^{\dagger}_{k, \lambda} G^{*}_{\beta \alpha }(k, \lambda) \Bigr ),
\end{equation}
\begin{equation}
G_{\alpha,\beta}(k, \lambda) = - \frac{e}{mc}\left (  \frac{2 \pi \hbar^{2} c}{k} \right)^{1/2} \mathbf{e}_{k, \lambda}
\langle \alpha| \exp \Bigl(\frac{i \mathbf{k} \mathbf{r}}{\hbar} \Bigr ) \cdot \mathbf{p} | \beta \rangle.
\end{equation}
Here $| \alpha \rangle$ and $| \beta \rangle$ are the eigenstates of energies  $E_{\alpha}$ and $E_{\beta}$ that of the Hamiltonian $H_{at}$, and
are given by
\begin{equation}
H_{at} | \alpha \rangle = E_{\alpha}| \alpha \rangle, \quad (\alpha, \beta) = (a, b).
\end{equation}
In the electric-dipole approximation we get
\begin{equation}
\varphi_{\alpha \beta} = - \frac{e}{mc} \langle \alpha| \mathbf{p} | \beta \rangle 
\sum_{k, \lambda} \left (  \frac{2 \pi \hbar^{2} c}{k} \right)^{1/2} \mathbf{e}_{k, \lambda} ( b_{k, \lambda}  + b^{\dagger}_{k, \lambda}).
\end{equation}
The matrix element of the dipole moment $\mathbf{d} = e \mathbf{r}$ between states $| \alpha \rangle$ and $| \beta \rangle$ is related
to the matrix element of the momentum $\mathbf{p}$ in the following way 
\begin{equation}
\langle \alpha| \mathbf{p} | \beta \rangle  =  - \frac{m}{e \hbar}( E_{\alpha} - E_{\beta} ) \mathbf{d}_{\alpha \beta},
\end{equation}
and we assume  that $\langle \alpha| \mathbf{p} | \alpha \rangle = 0.$\\
As was already mentioned, we use the Schr\"{o}dinger-type equation with damping for the quantity $\langle a_{ \alpha} \rangle$ which has the
form
\begin{equation}
\label{8.11}
  i\hbar \frac{d \langle a_{ \alpha}\rangle}{dt} = E_{\alpha}\langle a_{ \alpha} \rangle + \sum_{\beta }
K_{\alpha \beta} \langle a_{ \beta}\rangle,
\end{equation}
where
\begin{equation}
\label{8.12}
K_{\alpha \beta} = \frac{1}{i \hbar}\sum_{\gamma } \int_{- \infty}^{0} d t_{1}e^{\varepsilon t_{1}} 
\langle \varphi_{\alpha \gamma } \tilde{\varphi}_{\gamma \beta}(t_{1} )\rangle_{q}.
\end{equation}
Here $\tilde{\varphi}_{\alpha \beta}(t)$ is
\begin{equation}
\tilde{\varphi}_{\alpha \beta}(t) = \varphi_{\alpha \beta }(t) \exp \Bigl [\frac{i}{\hbar} (E_{\alpha} - E_{\beta})t \Bigr ].
\end{equation}
It is clear that the $K_{a a}$ and $K_{b a}$ are equal to zero and thus Eq.(\ref{8.11}) becomes
\begin{equation}
  i\hbar \frac{d \langle a_{b}\rangle}{dt} = E_{b}\langle a_{b}\rangle +  
K_{b b} \langle a_{b}\rangle,
\end{equation}
where
\begin{equation}
\label{8.15}
K_{b b} = \frac{2 \pi \hbar^{2} e^{2}}{m^{2} c} \frac{1}{\Omega} \sum_{k } \int_{- \infty}^{\infty} d \omega    \frac{1}{k}
\frac{J(k,\omega)}{\hbar \omega_{0} + \hbar \omega + i \varepsilon}A^{ab}_{ab}\left ( \frac{\mathbf{k}}{k}  \right).
\end{equation}
Here $\hbar \omega_{0} = (E_{b} - E_{a})$,
\begin{equation}\label{8.16}
 J(k,\omega) = \left( (\langle n_{k} \rangle + 1) \delta (\omega + c k)  
+  \langle n_{k} \rangle  \delta (\omega - c k) \right),
\end{equation}
\begin{equation}\label{8.17}
\langle n_{k} \rangle  =  \sum_{\lambda} \langle n_{k \lambda} \rangle  =\left ( e^{\beta c k}  - 1 \right)^{-1} = n(k),
\end{equation}
and
\begin{equation}\label{8.18}
 A^{ab}_{ab} =  | \langle a|\, \mathbf{p}\, | b \rangle |^{2}   -  \left( \langle a|\, \mathbf{p}\, | b \rangle  \frac{\mathbf{k}}{k}\right) 
 \left( \langle b|\, \mathbf{p} \, | a \rangle  \frac{\mathbf{k}}{k}\right).
\end{equation}
Next we have 
\begin{eqnarray}\label{8.19}
\frac{1}{\Omega} \sum_{k } \int_{- \infty}^{\infty} d \omega    \frac{1}{k}
\frac{J(k,\omega)}{\hbar \omega_{0} + \hbar \omega + i \varepsilon}  A^{ab}_{ab}\left ( \frac{\mathbf{k}}{k}  \right) = \\ \nonumber
 \frac{1}{(2\pi)^{3}} \int k d k \int_{- \infty}^{\infty} d \omega    \frac{1}{k}
\frac{J(k,\omega)}{\hbar \omega_{0} + \hbar \omega + i \varepsilon} \int A^{ab}_{ab}\left ( \frac{\mathbf{k}}{k}  \right) d \varpi,
\end{eqnarray}
where $d \varpi$ denotes the spherical angle element. It can be verified that
\begin{equation}
\label{8.20}
\int A^{ab}_{ab}\left ( \frac{\mathbf{k}}{k}  \right) d \varpi  =  \frac{8 \pi}{3}| \langle a|\, \mathbf{p}\, | b \rangle |^{2}.
\end{equation}
Substitution of Eq.(\ref{8.20}) into Eq.(\ref{8.15}) gives ($ \nu = c k$, $c$ is the speed of light)
\begin{eqnarray}
\label{8.21}
K_{b b} = \frac{2 e^{2}   }{m^{2} c^{2}  \hbar} |\langle a|\, \mathbf{p}\, | b \rangle |^{2}
\int_{0}^{\infty} \nu d \nu \left ( \frac{n(\nu) + 1}{\omega_{0} - \nu + i \varepsilon} + \frac{n(\nu)}{\omega_{0} + \nu + i \varepsilon}\right).
\end{eqnarray}
Finally, we obtain the formulae for width $\Gamma_{b}$, 
which we defined by $K_{b b} = \Delta E_{b} - (\hbar/2) i \Gamma_{b}$, from Eq.(\ref{8.21})
when the temperature tends to zero
\begin{equation}
\Gamma_{b}  =  \frac{4 }{3} \frac{e^{2} \omega_{0}}{m^{2} c^{3}  \hbar} | \langle a|\, \mathbf{p}\, | b \rangle |^{2} = 
\frac{4 }{3} \frac{ \omega_{0}^{3}}{ c^{3}  \hbar} |\mathbf{d}_{ab} |^{2}.
\end{equation}
This expression coincides with the well-known value for the natural width of spectral lines. We are not concerned with the calculation of 
the shift and discussion of its linear divergence because this is a usual example of the divergence of the self-energy in field theories. \\ Thus,
with the aid of the Schr\"{o}dinger-type equation with damping\index{Schrodinger-type equation with damping} one can simply 
calculate the energy width and shift. The Eq.(\ref{e7.15})   was used widely
in a number of concrete problems of line broadening due to perturbation~\cite{kuzem17}. 
%
%
\section{Generalized Van Hove Formula}
%
%
Microscopic descriptions of  dynamical behavior of condensed matter use the notion of correlations over space and time~\cite{zub74,kuzem17}.
Correlations over space and time in the density fluctuations of a fluid are responsible for the scattering of light when 
light passes through the fluid. The fluctuating properties are conveniently described in terms of time-dependent correlation functions
formed from the basic dynamical variables, e.g. the particle number density. The fluctuation-dissipation theorem~\cite{zub74,kuzem17}, shows that the susceptibilities
can be expressed in terms of the fluctuating properties of the system in equilibrium.
In  paper~\cite{kuzem12} the theory of scattering of particles (e.g. neutrons) by statistical medium was recast for the 
nonequilibrium statistical medium. The correlation scattering function of the relevant variables give rise to a very compact and entirely general expression for the
scattering cross section of interest. The formula obtained by Van Hove~\cite{kuzem17} provides a convenient method of 
analyzing the properties of slow neutron and light scattering by systems of particles such as gas, liquid or 
solid in the equilibrium state.
In the paper~\cite{kuzem12} the theory of scattering of particles by many-body system was  reformulated and generalized  for the 
case of nonequilibrium statistical medium. A  method of  quantum-statistical derivation the space and time Fourier 
transforms of the Van Hove correlation function
was   formulated on the basis of the method of NSO. 
This expression gives a natural extension of the familiar Van Hove formula for scattering of slow neutrons 
for the case in which the system under consideration is in a nonequilibrium state.\\
The differential cross section 
for the scattering of thermal neutrons may be expressed in terms of microscopic two-time correlation functions
of dynamical variables for the target system. For equilibrium systems, the van Hove formalism provides a general approach
to a compact treatment of scattering of neutrons (or other particles) by arbitrary systems of atoms 
in equilibrium~\cite{kuzem17}. 
The relation between the cross-sections for scattering of slow neutrons by an assembly of nuclei and space-time correlation functions for the
motion of the scattering system has been given by Van Hove in terms of the dynamic structure factor. Van Hove showed that the energy and angle differential cross
section is proportional to the double Fourier transform of a time-dependent correlation function $G(\textbf{r}, t)$. 
By definition, $G(\textbf{r}, t)$
is the equilibrium ensemble average of a product of  two time-dependent density operators and is therefore closely related to the linear
response of the system to an externally induced disturbance.\\
To formulate it a more precisely,  the dynamic structure factor\index{dynamic structure factor} is a mathematical function that contains information about 
inter-particle correlations and their time evolution. Experimentally, it can be accessed most directly by inelastic neutron 
scattering. The dynamic structure factor is most often denoted $\mathcal{S}(\textbf{k},\omega)$, where $\textbf{k}$   is a wave vector (a wave number for isotropic materials), 
and $\omega$ a frequency (sometimes stated as energy, $\hbar \omega$). It is the spatial and temporal Fourier transform of 
van Hove's time-dependent 
pair correlation function $G(\textbf{r},t)$, whose Fourier transform with respect to $\textbf{r}$, $\mathcal{S}(\textbf{k},t)$, is called the intermediate 
scattering function and can be measured by neutron spin echo spectroscopy.  In an isotropic sample (with scalar $r$), 
$G(r,t)$ is a time dependent radial distribution function.
In contrast with the systems in equilibrium state, no such general approach was formulated for the systems in
nonequilibrium state.\\
It is well known that the basic quantity is measured in the scattering experiment is the partial differential cross-section.
We will consider a target as a crystal with   lattice period $a$.
Transition amplitude is first order in the perturbation and the probability is consequently second order.
A perturbative approximation for the transition probability from an initial state to a final state under the action of a weak
potential $V$ is written as
\begin{equation}\label{e9.1}
 W_{kk'} = \frac{2 \pi}{\hbar } \left | \int d^{3}r   \psi^{*}_{k'} V \psi_{k} \right |^{2} D_{k'}(E'), 
\end{equation}
where $D_{k'}(E') $ is the density of final scattered states. Definition of the scattering cross-section is
\begin{equation}\label{e9.2}
 d \sigma  =   \frac{ W_{kk'}}{\textrm{Incident flux}}.  
\end{equation}
The incident flux is equal to $\hbar k' /m$ and the density of final scattered states is
\begin{equation}\label{e9.3}
  D_{k'}(E')  =   \frac{ 1}{(2 \pi)^{3}} \frac{d^{3}k' }{d E'} = \frac{m^{2}}{(2 \pi)^{3} \hbar^{3}}\, d \Omega \bigl (\frac{\hbar k'}{m} \bigr ).
\end{equation}
Thus, the differential scattering cross-section is written as
\begin{equation}\label{e9.4}
 \frac{d \sigma }{d \Omega }  =  
 \frac{m^{2}}{(2 \pi)^{2} \hbar^{4}}\, \frac{k'}{k} \left | \int d^{3}r  e^{ i (\mathbf{k}' - \mathbf{k}) \mathbf{r}} V (\mathbf{k})\right |^{2}.
\end{equation}
The general formalism described above can be applied to the particular case of neutron inelastic scattering~\cite{kuzem17,kuzem12}.
A typical experimental situation includes a monochromatic beam of neutrons, with energy $E$ and wave vector $\textbf{k}$, 
scattered by a sample or target.
Scattered neutrons are analyzed as a function of both their final energy $E' = E + \hbar \omega$, and the direction, $\vec{\Omega}$, of their final
wave vector, $\textbf{k}'$. We are interested in the quantity $I$, which is the number of neutrons scattered per second, 
between $\textbf{k}$ and $\textbf{k} + d \textbf{k}$
\begin{equation}\label{e9.5}
 I = I_{0} \frac{m a^{3}}{\hbar k}d w( \textbf{k} \rightarrow  \textbf{k}')D(\textbf{k})d \textbf{k}.
\end{equation}
Here, $m$ is the neutron mass, $a^{3}$ is the sample unit volume and $d w( \textbf{k} \rightarrow  \textbf{k}')$ is the transition probability from the initial
state $ |\textbf{k}\rangle$ to the final state $| \textbf{k}'\rangle$, and $D(\textbf{k})$ is the density of states of momentum $\textbf{k}$. It is given by
\begin{equation}\label{e9.6}
 D(\textbf{k})d \textbf{k} =   \frac{ a^{3}}{(2 \pi)^{3}} k^{2} d \vec{\Omega}  d k.
\end{equation}
It is convenient to take the following representations for the incident and scattered wave functions of a neutron:
\begin{equation}\label{e9.7}
 \psi_{k} = \sqrt{\frac{m}{k} }\, e^{\frac{i}{\hbar} (\textbf{k} \textbf{r})}, 
 \quad \psi_{k'} = \frac{1}{(2 \pi \hbar)^{3/2}} \,e^{ \frac{i}{\hbar} (\textbf{k}' \textbf{r})}.
\end{equation}
For the transition amplitude we obtain
\begin{equation}\label{e9.8}
d w( \textbf{k} \rightarrow  \textbf{k}') =   \frac{m}{\hbar^{2} k} \frac{d k'_{x}d k'_{y}d k'_{z}}{(2 \pi \hbar)^{3}} 
\int^{\infty}_{\infty}  \langle V (\textbf{r}) V(\textbf{r}', t) \rangle e^{[-i/\hbar(\textbf{k} - \textbf{k}')(\textbf{r} - \textbf{r}')  - i \omega t]}
dt d \textbf{r} d \textbf{r}'.
\end{equation}
In other words, the transition amplitude which describes the change of the state of the probe per unit time is
\begin{equation}\label{e9.9}
d w( \textbf{k} \rightarrow  \textbf{k}') =   \frac{1}{\hbar^{2}}  \int^{\infty}_{\infty} dt \textrm{Tr} \Bigl ( \rho_{m} V_{\textbf{k}'\textbf{k}}(0)V_{\textbf{k}' \textbf{k}}(t) \Bigr ) 
 \exp(- i \omega t), 
\end{equation}
where $\rho_{m}$ is a statistical matrix of the target.\\
Thus  the partial differential cross-section is written in the form
\begin{equation}\label{e9.10}
 \frac{d^{2} \sigma }{d \Omega d E'}  =  \frac{1 }{d \vec{\Omega} d \omega} \cdot \frac{I}{I_{0}}.
\end{equation}
It can be rewritten as
\begin{equation}\label{e9.11}
 \frac{d^{2} \sigma }{d \Omega d E'}  = A \int^{\infty}_{\infty}  \langle V (\textbf{r}) V(\textbf{r}', t) \rangle e^{[-i/\hbar(\textbf{k} - \textbf{k}')(\textbf{r} - \textbf{r}')  - i \omega t]}
dt d \textbf{r} d \textbf{r}',
\end{equation}
where
\begin{equation}\label{e9.12}
 A = \frac{m^{2} }{(2  \pi)^{3} \hbar^{5}}  \frac{k'}{k}, \quad   E' = \frac{k'^{2}}{2 m}.
\end{equation}
Thus the differential scattering cross section (in first Born approximation) for a system of interacting particles is 
written in the form (\ref{e9.11}),
where $A$ is a factor depending upon the momenta of the incoming and outgoing particles and upon the scattering potential for particle scattering, 
which for neutron scattering may be taken as the Fermi pseudopotential
\begin{equation}\label{e9.13}
 V = \frac{2  \pi \hbar^{2} }{m} \sum_{i}b_{i} \delta (\textbf{r} - \textbf{R}_{i}).
\end{equation}
Here $\textbf{R}_{i}$ is the position operator of nuclei in the target and $b_{i}$ is the corresponding scattering length.
It should be taken into account that
\begin{equation}\label{e9.14}
 V = \sum_{i=1}^{N} V (\textbf{r} - \textbf{R}_{i}) = \sum_{i=1}^{N} e^{- \frac{i}{\hbar} (\textbf{p} \textbf{R}_{i})} V (\textbf{r}) 
e^{\frac{i}{\hbar} (\textbf{p} \textbf{R}_{i})}, 
\end{equation}
and
\begin{equation}\label{e9.15}
\langle \beta \, \textbf{k}'|V|\alpha \, \textbf{k}\rangle  = \langle \textbf{k}'|V(\textbf{r})| \textbf{k}\rangle
\sum_{i=1}^{N} \langle \beta |e^{- \frac{i}{\hbar} (\textbf{k}' \textbf{R}_{i})} e^{ \frac{i}{\hbar} (\textbf{k} \textbf{R}_{i})}  |\alpha \rangle.
\end{equation}
Thus we obtain
\begin{equation}\label{e9.16}
 \frac{d^{2} \sigma }{d \Omega d E'}  \propto  \frac{k'}{k} \frac{1}{2 \pi} \sum_{ij} \int_{- \infty}^{\infty} \frac{1}{N}  b_{i}b_{j}
\Bigl \langle  \exp [ \frac{i}{\hbar} \vec{\kappa} \textbf{R}_{i}(0)]\, \exp [ - \frac{i}{\hbar} \vec{\kappa} \textbf{R}_{j}(t)] \Bigr \rangle
\exp ( - i \omega t) dt.
\end{equation}
The quantity 
measured in a neutron experiment is related to the imaginary (dissipative) part of the corresponding susceptibility. It is expressed as the weighted
sum of two susceptibilities: $\mathcal {S}_{c}(\kappa, \omega),$  which is called the coherent scattering law; 
and $\mathcal {S}_{ic}(\kappa, \omega),$  which is called incoherent (or single-particle) scattering law. Here $\kappa = k - k'.$
The self-correlation function $G_{ic}(\textbf{r}, t)$, introduced by Van hove, was widely used in the analysis of the incoherent scattering
of slow neutrons by system of atoms of molecules; it appears also in calculations of the line shapes for resonance absorption of neutrons
and gamma rays. The correlation function approach is of particular utility when the scattering system is in a nonequilibrium state: a dense gas, liquid or crystal, in which the
dynamics of atomic motions are very complex. Thus, it is clear that a more sophisticated theoretical approach 
to the problem should be elaborated. To be able to describe the effects of retardation and dissipation properly we
will proceed in a direct analogy with the derivation of the kinetic equations for a system in a thermal bath~\cite{kuzem07,wkuz70}, 
discussed above.\\ 
We consider  a statistical medium (target) with Hamiltonian $H_{m}$, a probe (beam) with Hamiltonian $H_{b}$ and an 
interaction $V$ between the two $H = H_{0} + V = H_{m} + H_{b} + V. $ 
We will consider  statistical medium in a  nonequilibrium state.
Consider the expression for
the transition amplitude which describes the change of the state of the probe per unit time  
\begin{equation}\label{e9.17}
d w( \textbf{k} \rightarrow  \textbf{k}') =   \frac{1}{\hbar^{2}}  \int^{\infty}_{\infty} 
dt \textrm{Tr}_{m} \Bigl ( \rho_{m}(t) V_{\textbf{k}'\textbf{k}}(0)V_{\textbf{k}' \textbf{k}}(t)\Bigr ) 
 \exp(- i \omega t), 
\end{equation}
where $\rho_{m}(t)$ is the nonequilibrium statistical operator of  target. 
Thus  the partial differential cross-section is written in the form
\begin{equation}\label{e9.18}
 \frac{d^{2} \sigma }{d \Omega d E'}  = A \int^{\infty}_{\infty}  \langle V (\textbf{r}) V(\textbf{r}', t) \rangle_{m} e^{[-i/\hbar(\textbf{k} - \textbf{k}')(\textbf{r} - \textbf{r}')  - i \omega t]}
dt d \textbf{r} d \textbf{r}',
\end{equation}
where
\begin{equation}\label{e9.19}
 A = \frac{m^{2} }{(2  \pi)^{3} \hbar^{5}}  \frac{k'}{k}, \quad   E' = \frac{k'^{2}}{2 m}.
\end{equation}
and
$\langle \ldots \rangle_{m} = \textrm{Tr}_{m}(\rho^{m}(t) \ldots).$
Again, we took into account that
\begin{equation}\label{e9.20}
\langle \alpha' \, \textbf{k}'|V|\alpha \, \textbf{k}\rangle  = \langle \textbf{k}'|V(\textbf{r})| \textbf{k}\rangle
\sum_{i=1}^{N} \langle \alpha' |e^{- \frac{i}{\hbar} (\textbf{k}' \textbf{R}_{i})} e^{ \frac{i}{\hbar} (\textbf{k} \textbf{R}_{i})}  |\alpha \rangle.
\end{equation}
Thus we obtain
\begin{eqnarray}\label{e9.21}
 \frac{d^{2} \sigma }{d \Omega d E'}  =  \qquad \\ \nonumber \frac{- 1}{(i \hbar)^{2}} \tilde{A} \sum_{i,j=1}^{N}\int_{0}^{t} d \tau
\sum_{\alpha} \langle \alpha | 
\{ \exp [ \frac{i}{\hbar} \vec{\kappa} \textbf{R}_{i}(\tau - t)]  \exp [ \frac{i}{\hbar} \vec{\kappa} \textbf{R}_{j}(0)] \exp (i \omega (\tau - t))\\  
+ \exp [ \frac{i}{\hbar} \vec{\kappa} \textbf{R}_{i}(0)] \exp [ \frac{i}{\hbar} \vec{\kappa} \textbf{R}_{j}(\tau - t)]
\exp (- i \omega (\tau - t))
\} \rho^{m}(t) | \alpha \rangle. \nonumber
\end{eqnarray}
It can be rewritten in another form~\cite{kuzem17,kuzem12}
\begin{eqnarray}\label{e9.22}
 \frac{d^{2} \sigma }{d \Omega d E'}  =  \qquad \\ \nonumber \frac{- 1}{(i \hbar)^{2}} \tilde{A} \sum_{i,j=1}^{N}\int_{0}^{t} d \tau
\{2 \textrm{Re}  \left  \langle  
 \exp [ \frac{i}{\hbar} \vec{\kappa} \textbf{R}_{i}(\tau - t)]  \exp [ \frac{i}{\hbar} \vec{\kappa} \textbf{R}_{j}(0)] \right \rangle_{m}  \exp (i \omega (\tau - t)). \\  
\nonumber
\end{eqnarray}
In terms of the density operators $n_{ \vec{\kappa}} = \sum_{i}^{N} \exp \left (  i \vec{\kappa} \textbf{R}_{i} / \hbar  \right )$ the differential cross section take the form
\begin{equation}\label{e9.23} 
\frac{d^{2} \sigma }{d \Omega d E'}  = \tilde{A} 2 \textrm{Re} \mathcal{S}( \vec{\kappa}, \omega, t),
\end{equation}
where
\begin{equation}\label{e9.24}
 \mathcal{S}( \vec{\kappa}, \omega, t) =   \frac{- 1}{(i \hbar)^{2}}   \int_{0}^{t} d \tau \exp [  i \omega(\tau - t)] 
\langle  n_{ \vec{\kappa}} (\tau - t) n_{ - \vec{\kappa}}  \rangle_{m},
\end{equation}
Let us construct the nonequilibrium statistical operator of the medium. To do so, we should follow the
basic formalism of the nonequilibrium statistical operator method. According to this approach
we should take into account that
\begin{eqnarray}\label{e9.25}
\rho^{m} = \overline {\rho_{q}(t,0)} = \varepsilon \int^{0}_{-\infty} d \tau 
e^{\varepsilon \tau} \rho_{q}(t + \tau, \tau)\nonumber =  \\  \varepsilon \int^{0}_{-\infty} d \tau 
e^{\varepsilon \tau} \exp \left(- \frac{H_{m} \tau}{i\hbar} \right )  \rho_{q}(t + \tau, 0) \exp \left( \frac{H_{m} \tau}{i\hbar} \right )  =   
\varepsilon \int^{0}_{-\infty} d \tau e^{\varepsilon \tau} \exp \left(- S(t + \tau, \tau) \right ). \quad
\end{eqnarray} 
Thus the nonequilibrium statistical operator of the medium will take the form
\begin{eqnarray}\label{e9.26}
\rho^{m} (t,0) = \exp \left(- S(t, 0) \right )  \qquad \\ +  \int^{0}_{-\infty} d \tau e^{\varepsilon \tau} \int^{1}_{0} d \tau' 
\exp \left(- \tau' S(t + \tau, \tau) \right ) \dot{S}(t + \tau, \tau) \exp \left(- (\tau'-1) S(t + \tau, \tau) \right ),\nonumber 
\end{eqnarray}
where
\begin{eqnarray}\label{e9.27}
\dot{S}(t, \tau) = \exp \left(- \frac{H_{m} \tau}{i\hbar} \right ) \dot{S}(t, 0) \exp \left( \frac{H_{m} \tau}{i\hbar} \right )
\end{eqnarray}
and
\begin{eqnarray}\label{e9.28}
\dot{S}(t, 0) = \frac{\partial S (t, 0) }{\partial t} + \frac{1}{i \hbar}[ S (t, 0), H ] = \\ \nonumber
\sum_{m}  \left(  \dot{P}_{m} F_{m}(t)  +  (P_{m} -\langle \dot{P}_{m} \rangle^{t}_{q})  \dot{F}_{m}(t) \right).
\end{eqnarray}
Finally, the general expression for the scattering function of beam of neutrons by the nonequilibrium medium in the
approach of the nonequilibrium statistical operator method is given by
\begin{eqnarray}\label{e9.29}
 \mathcal{S}( \vec{\kappa}, \omega, t) =    \frac{- 1}{(i \hbar)^{2}}   \int_{0}^{t} d \tau  
 \langle  n_{ \vec{\kappa}} (\tau - t) n_{ - \vec{\kappa}}(0)  \rangle_{q}^{t}
 \exp [  i \omega(\tau - t)]   \quad \\ \nonumber + \frac{- 1}{(i \hbar)^{2}}   \int_{0}^{t} d \tau   \int_{- \infty}^{0} d \tau' e^{\varepsilon \tau'}
\Bigl (
n_{ \vec{\kappa}} (\tau - t) n_{ - \vec{\kappa}}(0), \dot{S}(t + \tau')  \Bigr )^{t + \tau'} \exp [  i \omega(\tau - t)].  
\end{eqnarray}
Here the standard notation~\cite{zub74}  for $(A,B)^{t}$ were introduced
\begin{equation}\label{e9.30}  
(A,B)^{t} = \int_{0}^{1} d \tau \textrm{Tr}  \left [A \exp ( -\tau S (t, 0) ) (B -\langle B \rangle^{t}_{q})   \exp ( (\tau - 1)  S (t, 0) ) \right ],
\end{equation}
\begin{equation}\label{e9.31} 
\rho_{q}(t, 0) = \exp ( -  S (t, 0) ); \quad \langle B \rangle^{t}_{q} = \textrm{Tr} (B \rho_{q}(t, 0)).
\end{equation}
Now we show that the problem of finding of the nonequilibrium statistical operator for the beam of neutrons
has many common features with the description of the small subsystem interacting with thermal reservoir.\\ 
Let us consider again the our Hamiltonian. The state of the overall system at time $t$ is given by the statistical operator
\begin{equation}\label{e9.32}
 \rho(t) = \exp \left ( \frac{- i H_{0} t}{\hbar}\right ) \rho(0) \exp \left ( \frac{ i H_{0} t}{\hbar}\right ), \quad  \rho(0) = \rho^{m}(0) \otimes \rho^{b} (0). 
\end{equation}
The initial state
assumes a factorized form $(\rho^{m}(0) $ and $\rho^{b}(0) $   correspond  to the density operators that represent the initial 
states of the system and the probe, respectively). The state of the system and the probe at time $t$ can be described 
by the reduced density operators
\begin{eqnarray}\label{e9.33}
  \rho^{b} (t) = \textrm{Tr}_{m} [ \rho(t)]  = \textrm{Tr}_{m}\left ( \exp ( \frac{- i H_{0} t}{\hbar}) \rho^{m}(0) \otimes \rho^{b} (0) \exp ( \frac{ i H_{0} t}{\hbar})\right ),
 \\
\label{e9.34}   
\rho^{m} (t) = \textrm{Tr}_{b} [ \rho(t)]  = \textrm{Tr}_{b}\left ( \exp ( \frac{- i H_{0} t}{\hbar}) \rho^{m}(0) \otimes \rho^{b} (0) \exp ( \frac{ i H_{0} t}{\hbar})  \right ),    
\end{eqnarray}
where $ \textrm{Tr}_{m}$ and $ \textrm{Tr}_{b}$ stands for a partial trace over the system (statistical medium) and the 
beam (probe) degrees of freedom, respectively.\\
In quantum theory a transition probability from a state of a statistical system which is described by density matrix $\rho_{i}$ to the state
$\rho_{f}$ ("i" - initial, "f" -final)   is given by
\begin{equation}\label{e9.35}
W_{if} (t)  =  \textrm{Tr}(\rho^{i}(t)\rho^{f}(t) ). 
\end{equation}
It is reasonable to assume that $\rho^{i}$ has the form $ \rho^{i}(t)   = \rho^{i}(0) = |k \rangle \langle k |$. Then the transition probability per unit time
takes the form
\begin{eqnarray}\label{e9.36}
w_{if} (t)  =  \frac{d}{dt}   \textrm{Tr}(|k \rangle \langle k | \rho^{f}(t) ) =  \nonumber \frac{d}{dt} \langle k | \rho^{f}(t)|k \rangle =
\langle k |\frac{d}{dt} \rho^{f}(t)|k \rangle.
\end{eqnarray}
Let us consider an extended Liouville equation for the statistical medium (target) with Hamiltonian $H_{m}$, a probe (beam) with Hamiltonian $H_{b}$ and an interaction $V$ between the two.
The density matrix $\rho(t)$ for the combined medium-beam complex obeys
\begin{equation}\label{e9.37}
 \frac{\partial}{\partial t}\rho(t) - \frac{1}{i\hbar} [(H_{m} + H_{b} + V), \, \rho(t)]_{-} = - \varepsilon \left ( \rho(t) - P \rho(t) \right ).
\end{equation}
Here $P$ is projection superoperator   with the properties:
$$ P^{2} = P, \, P ( 1 - P) = 0,\, P ( A + B) = P A + P B.$$ 
The simplest possibility is
\begin{equation}\label{e9.38}
P \rho(t)  =  \rho^{m0} \rho^{b} = \rho^{m0} \sum_{\alpha}  \langle \alpha | \rho(t) |\alpha \rangle. 
\end{equation}
Here  $\rho^{m0}$ is the equilibrium  statistical operator of the medium.\\ 
It will be reasonable to adopt for the nonequilibrium medium   the following boundary condition
\begin{equation}\label{e9.39}
 \frac{\partial}{\partial t}\rho(t) - \frac{1}{i\hbar} [(H_{m} + H_{b} + V), \, \rho(t)]_{-} = 
- \varepsilon \left ( \rho(t) -   \rho^{m}(t) \rho^{b}(t) \right ),
\end{equation}
where
\begin{equation}\label{e9.40}  
\rho^{m}(t) = \textrm{Tr}_{b}(\rho(t)) =  \sum_{k}  \langle k | \rho (t) |k \rangle,  
\end{equation}
and (in general case)
\begin{eqnarray}\label{e9.41} 
\rho^{b} = \textrm{Tr}_{m}(\rho(t)) =  \sum_{\alpha}  \langle \alpha | \rho (t) |\alpha \rangle   =
\sum_{k k'}  \langle k' |\rho_{b}(t)|k \rangle |k \rangle \langle k' | = 
\sum_{k k'} \rho^{b}_{k'k } |k \rangle \langle k' |. 
\end{eqnarray}
Thus, according to the nonequilibrium statistical operator method, we can rewrite 
Eq.(\ref{e9.37}) in the form
\begin{equation}\label{e9.42}
 \frac{\partial}{\partial t}\rho(t) - \frac{1}{i\hbar} [H, \, \rho(t)]_{-} = 
- \varepsilon \left ( \rho(t) -   \rho^{m}(t) \sum_{q} \rho^{b}_{qq} (t)|q \rangle \langle q | \right ),
\end{equation}
where we confined ourselves to the $\rho^{b}$ diagonal in states $|q \rangle$ and
$(\varepsilon \rightarrow 0)$ after the thermodynamic limit. 
The required nonequilibrium statistical operator in accordance with Eq.(\ref{e9.42}) is defined as 
\begin{eqnarray}\label{e9.43}
\rho_{\varepsilon}  = \rho_{\varepsilon}(t,0)   =  \overline {\rho_{q}(t,0)} = 
\varepsilon \int^{0}_{-\infty} d \tau 
e^{\varepsilon \tau} \rho_{q}(t + \tau, \tau) = \\ \varepsilon \int^{0}_{-\infty} d \tau U(\tau)
\rho^{m}(t) \sum_{k} \rho^{b}_{kk} (t + \tau)|k \rangle \langle k |U^{\dag}(\tau). \nonumber
\end{eqnarray}
Here $U(t)$ is the operator of evolution.\\
In direct analogy with the derivation of the evolution equation  for the small subsystem (beam) interacting with thermal
reservoir (noequilibrium statistical medium) we find
\begin{eqnarray}\label{e9.44}
\frac{\partial}{\partial t}\rho^{b}_{qq}(t) =   - \frac{1}{i\hbar} \varepsilon \int^{0}_{-\infty} d \tau e^{\varepsilon \tau} \\  \times \sum_{k}
\rho^{b}_{kk} (t + \tau) \sum_{\alpha}  
\langle \alpha | \langle q |\left [U(\tau)\rho^{m}(t) |k \rangle \langle k | U^{\dag}(\tau), V \right ]_{-} |q \rangle | \alpha \rangle.
\nonumber
\end{eqnarray}
In  analogy with the derivation of the evolution equation  for the small subsystem,
in the lowest order approximation it is reasonably to consider that $\rho^{b}_{kk} (t + \tau) \simeq \rho^{b}_{kk} (t).$
This approximation means neglecting to the memory effects.
After integration by parts we obtain the evolution equation of the form
\begin{eqnarray}\label{e9.45}
\frac{\partial}{\partial t}\rho^{b}_{qq}(t) =    \frac{1}{\hbar^{2}} \sum_{k} \rho^{b}_{kk} (t) \\  \times \int^{0}_{-\infty} d \tau  e^{\varepsilon \tau}
 \sum_{\alpha}  
\langle \alpha | \langle q |\left  [U(\tau)[V(\tau),  \rho_{m}(t) |k \rangle \langle k| ]_{-} \, U^{\dag}(\tau), V \right ]_{-} |q \rangle |\alpha \rangle.
\nonumber
\end{eqnarray}
As before, we confined ourselves to the second order in the perturbation $V$. This assumption gives also that in Eq.(\ref{e9.45}) 
$U = U^{\dag} = 1.$ As a result we arrive to the equation similar in the form to the equation Eq.(\ref{e6.11})
\begin{equation}\label{e9.46}
\frac{\partial}{\partial t}\rho^{b}_{kk}(t) =    \sum_{q} W_{q \rightarrow k}  \rho^{b}_{qq}(t)
 - \sum_{q} W_{k \rightarrow q}  \rho^{b}_{kk}(t).
 \end{equation}
The explicit expression for the "\emph{effective transition probabilities}" $W_{q \rightarrow k}$ is given by the formula
\begin{eqnarray}\label{e9.47}
W_{q \rightarrow k} = 2 \textrm{Re} \frac{1}{\hbar^{2}} \int^{0}_{-\infty} d \tau  e^{\varepsilon \tau}
\left \langle V_{q k} V_{k q} (\tau) \right \rangle^{t}_{m}.
\end{eqnarray}
Here $V_{q k} = \langle q| V|k \rangle,$   $\langle \ldots \rangle^{t}_{m} = \textrm{Tr} (\ldots  \rho^{m}(t))$
and $(\varepsilon \rightarrow 0)$ after the thermodynamic limit. Thus we generalized the 
expressions (\ref{e9.1}),(\ref{e9.9})  and (\ref{e9.17}) for the nonequilibrium media. That lead to a straightforward
foundation of formula (\ref{e9.24})
and the problem of the derivation of the Van Hove formula for  scattering of neutrons on a nonequilibrium
statistical medium is completed.  
%
%
%
%
%
\section{Concluding Remarks}
%
In this paper we analyzed the foundational issues and some applications of the   method of nonequilibrium 
statistical operator, formulated by Zubarev~\cite{zub74}, in connection with some other approaches. 
By contrasting it with other methods, we tried to stress the innovative character  of the NSO formalism and its 
internal consistency and operational ability for solving concrete problems.\\
To elucidate the nature of transport and relaxation processes, 
the generalized kinetic equations  were described also for a system weakly coupled to a thermal bath~\cite{kuzem05,wkuz70,wzkuz70}.
It was shown that the "collision term" has the same characteristic functional form as for the generalized kinetic
equations for the system with small interactions among  particles.
The applicability of the general formalism to physically relevant situations was investigated.
It was shown that some 
known generalized  kinetic equations  naturally  emerges  within the NSO formalism~\cite{kuzem07,kuzem05,kuzem11}. 
Relaxation of a small dynamic  subsystem in contact with a thermal bath was considered on the basis of the
derived equations. 
It is of especial interest that
\emph{the Schr\"{o}dinger-type equation} for the average amplitude  describing the energy shift and damping of a
particle in a thermal bath  and the coupled kinetic equation  describing the dynamic  and statistical
aspects of the motion  were obtained as well by this method. The equations derived  can help in the understanding 
of quantum evolution and of the origin of irreversible behavior in quantum phenomena.\\  
Additional material and discussion of these and related problems can be found 
in Refs.~\cite{kuzem17,kuzem07,kuzem11,kuzem16,kuzrnc18,kuzem18,kuzem12} 

\begin{thebibliography}{99}\itemsep -1mm
%
%
%
\bibitem{bogol46}
N. N. Bogoliubov, \emph{Problems of Dynamical Theory in Statistical Physics.}  
Gostekhizdat, Moscow  (1946) [in Russian].
%
%
%
\bibitem{bog62} N. N. Bogoliubov, Problems of Dynamical Theory in Statistical Physics.
in:  {\em Studies in  Statistical Mechanics},
 J. de  Boer and G. E. Uhlenbeck,  (eds.),
North-Holland, Amsterdam   (1962), vol. 1, p.1-118.
%
%
%
\bibitem{nnb78}
N. N. Bogoliubov, On the stochastic processes in the dynamical systems. 
\emph{Sov. J. Part. Nucl.,}  \textbf{9}, 205  (1978). 
%
%
\bibitem{dpetr09}
D. Ya. Petrina, {\em Stochastic Dynamics and Boltzmann Hierarchy.} 
Walter de Gruyter,   Berlin, New York  (2009).
%
%
%
\bibitem{zub74} D. N. Zubarev,
{\em Nonequilibrium Statistical Thermodynamics.}  Consultant Bureau,   New York   (1974).
%
%
%
\bibitem{macl89} J. A. McLennan,
{\em Introduction to Nonequilibrium Statistical Mechanics.} Prentice Hall,  New Jersey  (1989).
%
%
\bibitem{eu98}
Byung Chan Eu,  {\em Nonequilibrium Statistical Mechanics. Ensemble Method.} 
Kluwer Publ., Boston, London  (1998).
%
%
%
\bibitem{zwa65}  R. Zwanzig, Time-correlation functions and transport coefficients in
statistical mechanics,  in: {\it Annual Review of Physical Chemistry,}
John Wiley and Sons, New York  (1965), vol. 16, p.~67.
%
%
%
\bibitem{zw70}
R. Zwanzig, The concept of irreversibility in statistical mechanics.  
{\it Pure and Appl. Chem.,} \textbf{22}, 371 (1970).
%
%
%
\bibitem{rbal97}
R. Balescu,
\emph{Statistical Dynamics: Matter Out of Equilibrium.} 
World Scientific, Singapore  (1997).
%
%
%
\bibitem{zwan01} R. Zwanzig,
{\em Nonequilibrium Statistical Mechanics.}   Oxford University Press, Oxford   (2001).
%
%
%
\bibitem{galla14} G. Gallavotti,
{\em   Nonequilibrium and Irreversibility,}  Springer, Berlin   (2014).
%
%
\bibitem{kozlov08}
V. V. Kozlov, \emph{Gibbs Ensemblies and  Nonequilibrium Statistical Mechanics.}  
Regular and Chaotic Dynamics, Moscow  (2008) [in Russian].
%
%
%
\bibitem{vede11}
V. Vedenyapin, A.Sinitsyn,  E. Dulov, \emph{Kinetic Boltzmann, Vlasov and Related Equations}.
Elsevier, Amsterdam  (2011).  
%
%
\bibitem{kuzem17}  A. L. Kuzemsky,
\textit{Statistical Mechanics and the Physics of Many-Particle Model  Systems}. 
World Scientific, Singapore  (2017).
%
%
%
\bibitem{kuzem07}
A.L. Kuzemsky, Theory of transport processes and the method of the nonequilibrium
statistical operator. \emph{Int. J. Mod. Phys.,} B \textbf{21},  2821-2949 (2007).
%
%
%
\bibitem{kuzem05}
A. L. Kuzemsky, Generalized kinetic and evolution equations in the approach of the nonequilibrium statistical operator. 
{\em Int. J. Mod. Phys.,} B \textbf{19}, 1029-1059 (2005).  
%
%
\bibitem{kuzem11}
A.L. Kuzemsky, Electronic transport in metallic systems  and generalized kinetic equations.
\emph{Int. J. Mod. Phys.,} B \textbf{25}, 3071-3183  (2011).  
%
%
%
\bibitem{gib1} 
J. W. Gibbs, {\em Elementary Principles in Statistical Mechanics Developed with Especial Reference to the Rational
Foundations of Thermodynamics}. Dover Publ., New York  (1960). 
%
%
%
\bibitem{kozlov00}
V. V. Kozlov,  
Thermodynamics of Hamiltonian Systems and Gibbs Distribution.
{Doklady Mathematics} \textbf{61}, 123-125,  (2000).
%
%
%
\bibitem{koz02}
V. V. Kozlov,
On Justification of Gibbs Distribution.
\textit{Regular and Chaotic Dynamics}  \textbf{7},  1-10.  (2002).
%
%
%
\bibitem{mehra98}
J. Mehra,  Josiah Willard Gibbs and the foundations of statistical mechanics.
{\em Found. Phys.}  \textbf{28}, 1785  (1998). 
%
%
%
\bibitem{kuzem14}
A. L. Kuzemsky,  Thermodynamic limit in statistical physics. 
{\em Int. J. Mod. Phys.,} B  \textbf{28}, 1430004 (28 pages) (2014).
%
%
%
\bibitem{minl}
R. A. Minlos,
\emph{Introduction to Mathematical Statistical Physics.} (University Lecture Series)  
American Mathematical Society  (2000).
%
%
%
\bibitem{jayn03}
E. T. Jaynes,  \emph{Probability Theory: The Logic of Science.}   
Cambridge University  Press, New York  (2003). 
%
%
%
\bibitem{koz06}
 V. V. Kozlov, O. G. Smolyanov, 
Information entropy in problems of classical and quantum statistical mechanics.
\emph{Doklady Mathematics},   \textbf{74}, 910  (2006).
%
%
%
\bibitem{kuzem16}
A. L. Kuzemsky, Probability, information and   statistical physics. 
\emph{Int. J. Theor. Phys.,}    \textbf{55}, 1378-1404  (2016).  
%
%
%
\bibitem{kuzrnc18}
A. L. Kuzemsky, Temporal evolution, directionality  of time and irreversibility. 
\textit{Rivista del Nuovo Cimento} \textbf{41},  513--574  (2018).
%
%
%
\bibitem{ross08}
 J. Ross,   \emph{Thermodynamics and Fluctuations far from Equilibrium}.
Springer, Berlin   (2008). 
%
%
%
\bibitem{scha14}
G. Schaller,  
\emph{Open Quantum Systems Far from Equilibrium}.  Springer, Berlin (2014). 
%
%
%
\bibitem{tome15}
T. Tome,  M.J. de Oliveira,
 \emph{Stochastic Dynamics and Irreversibility}.   Springer, Berlin (2015).
%
%
%
%
\bibitem{demi14} Y. Demirel,
{\em Nonequilibrium Thermodynamics: Transport and Rate Processes in Physical, Chemical 
and Biological Systems.}  Elsevier,   Amsterdam    (2014).
%
%
%
\bibitem{msuz11}
M. Suzuki, Irreversibility and entropy production in transport phenomena.
\emph{Physica,} A \textbf{390}, 1904  (2011).  
%
%
%
\bibitem{msuz11a}
M. Suzuki, First-principle derivation of entropy production in transport phenomena.
\emph{J. Phys: Conf. Ser.,}  \textbf{297}, 012019  (2011).  
%
%
%
\bibitem{msuz12}
M. Suzuki, Irreversibility and entropy production in transport phenomena, II: 
Statistical-mechanical theory on steady states including thermal disturbance and energy supply.
\emph{Physica,} A \textbf{391}, 1074  (2012).  
%
%
%
\bibitem{zwa60}
R. Zwanzig, Ensemble method in the theory of irreversibility. 
\emph{J. Chem. Phys.,}  \textbf{33}, 1338 (1960). 
%
%
%
\bibitem{berg55} P. G. Bergmann, J. L. Lebowitz,  New approach to nonequilibrium processes. 
{\em Phys. Rev.} \textbf{99}, 578 (1955).
%
%
%
\bibitem{leb57} J. L. Lebowitz, P. G. Bergmann,  Irreversible Gibbsian ensembles. 
{\em Ann. Phys.,} (N.Y.) \textbf{1}, 1 (1957).
%
%
\bibitem{leb59} J. L. Lebowitz, Stationary nonequilibrium Gibbsian ensembles.  
{\em Phys. Rev.} \textbf{114}, 1192 (1959).
%
%
\bibitem{leb62} J. L. Lebowitz, A. Shimony, Statistical mechanics of open systems. 
{\em Phys. Rev.} \textbf{128}, 1945 (1962).
%
%
%
\bibitem{cawe}
H. B. Callen and T. A. Welton, Irreversibility and generalized noise. 
\emph{Phys. Rev.} \textbf{83}, 34 (1951).   
%
%
\bibitem{mayer61} J. E. Mayer, Approach to thermodynamic equilibrium.
\emph{J. Chem. Phys.} \textbf{34}, 1207-1223   (1961).
%
%
%
\bibitem{bmitr}
N. N. Bogoliubov, Yu. A. Mitropolsky, 
{\em Asymptotical Methods in the Theory of Nonlinear Oscillations.}  
Gordon and Breach, New York  (1961).
%
%
%
\bibitem{mitr2}
Yu. A. Mitropolsky, {\em Averaging Method in Nonlinear Mechanics.}  
Naukova Dumka, Kiev  (1971)  [in Russian]. 
%
%
%
\bibitem{sam94}
A. M. Samoilenko, N. N. Bogoliubov and nonlinear mechanics, 
{\em  Uspekhi. Mat. Nauk,} 49, 103 (1994).
%
%
%
\bibitem{arnold} V. I. Arnold, From averaging to statistical physics. in: \emph{Problems of Modern Mathematical Physics},
Trudy Math. Inst. Steklov, Nauka, Moscow  (2000), \textbf{228}, p.196-202.
%
%
%
\bibitem{bog82} N. N. Bogoliubov, On some problems connected with the foundations of statistical 
mechanics.
in:  {\em Proc. Int. Symp. on Selected Topics  in  Statistical Mechanics},  
N. N. Bogoliubov, Jr., \emph{et al}., (eds.), JINR, Dubna, (1982),  p.9--18.
%
%
%
\bibitem{bpk69} 
N. N. Bogoliubov,  D. Ya. Petrina   and B. I. Khatset, Mathematical description of equilibrium state of classical 
systems on the basis of canonical formalism. \emph{Teor. Mat. Fiz.}  \textbf{1},  251 (1969).
%
%
%
\bibitem{earm96} J. Earman,  M. Redei,
Why ergodic theory does not explain the success of equilibrium statistical mechanics.
{\em Philosophy of Science,}   \textbf{47}, 63 (1996). 
%
%
%
\bibitem{kozl03}
V. V. Kozlov, D. V. Treschev,  
On New Forms of the Ergodic Theorem.
{\em  J. Dynam. Control Syst.}  \textbf{9}, no.~3,  449-453 (2003).  
%
%
%
\bibitem{bogach07}
V. I. Bogachev,   A. V.  Korolev,  
On the Ergodic Theorem in the Kozlov-Treshchev Form.
{\em Doklady Mathematics}, \textbf{75}, no.~1,  47-52  (2007). 
%
%
%
\bibitem{dnz55}
N. N. Bogoliubov, D. N. Zubarev, Method of asymptotic approximation and its application to the
motion of charged particles in the magnetic field. {\em  Ukrainian Math. J.,} \textbf{7}, 5 (1955).
%
%
%
\bibitem{espo10}
M. Esposito, K. Lindenberg,    C. Van den Broeck,
Entropy production as correlation between system and reservoir.
 {\em New J. Phys.}  \textbf{12}, 013013  (2010).
%
%
%
\bibitem{mart06}
L. M. Martyushev, V. D. Seleznev, Maximum entropy production principle in physics, chemistry and biology. 
{\it Phys. Rep.}   \textbf{426},  1 (2006).
%
%
%
\bibitem{galga72}
L. Galgani, A.Scotti, Planck-like distributions in classical and nonlinear mechanics.
{\it Phys. Rev. Lett.}   \textbf{28},  1173-1176 (1972).
%
%
%
\bibitem{lgalga00}
L. Galgani, A.Carati, Planck's formula and glassy behavior in classical nonequilibrium statistical
mechanics. {\it Physica}  A \textbf{280},  106--114 (2000).
%
%
\bibitem{galg01}
L. Galgani, A.Carati,  The theory of dynamical systems and the relations between
classical and quantum mechanics. \emph{Found. Phys.} \textbf{31}, 69--87  (2001). 
%
%
\bibitem{carati06}
A.Carati, L. Galgani,  A. Giorgilli,    Dynamical systems and thermodynamics,
in: \emph{Encyclopedia of Mathematical Physics}. Academic Press  (2006), p.125. 
%
%
%
\bibitem{zub69}
D. N. Zubarev, V. P. Kalashnikov, Extremal properties of the nonequilibrium statistical operator. 
{\em Teor. Mat. Fiz.,}  \textbf{1}, 137 (1969).
%
%
\bibitem{zub70}
D. N. Zubarev, V. P. Kalashnikov, Construction of statistical operators for nonequilibrium processes. 
{\em Teor. Mat. Fiz.,}  \textbf{3}, 126 (1970).
%
%
\bibitem{zuk70}
D. N. Zubarev, V. P. Kalashnikov, Derivation of the nonequilibrium statistical operator from the extremum of the 
information entropy.  {\em Physica,}  \textbf{46}, 550 (1970).
%
%
%
\bibitem{qbog} N. N. Bogoliubov, Quasiaverages in  problems of statistical mechanics. in:
\emph{Statistical Physics and Quantum Field Theory,} Nauka, Moscow, (1973), p. 7 [in Russian].
%
%
%
\bibitem{zubar70}
D. N. Zubarev, Boundary conditions for statistical operators in the theory of nonequilibrium processes 
and quasiaverages. {\em Teor. Mat. Fiz.,} \textbf{3}, 276 (1970).
%
%
%
%
\bibitem{kuzem10}
A. L. Kuzemsky,  Bogoliubov's vision: quasiaverages and broken symmetry  to quantum protectorate and emergence.
\emph{Int. J. Mod. Phys.}  B  \textbf{24},  835--935  (2010).
%
%
%
\bibitem{aus84}
M. I. Auslender, V. P. Kalashnikov, Equivalence of two forms of the nonequilibrium statistical 
operator. {\em Teor. Mat. Fiz.,}  \textbf{58}, 299 (1984).
%
%
%
\bibitem{aus03}
M. I. Auslender, Equivalence of two nonequilibrium ensembles based on maximum entropy principle.
arXiv:0308379[cond-mat.], (2003).
%
%
%
\bibitem{kuzem18}
A. L. Kuzemsky, Nonequilibrium Statistical Operator Method and Generalized Kinetic 
Equations. \emph{Theor. Math. Phys.} \textbf{194}, no.~1,  30--56. (2018).
%
%
%
\bibitem{pokrov2}
L. A. Pokrovski, Derivation of generalized kinetic equations with the aid of the method of the nonequilibrium statistical operator.
{\it Doklady Acad. Nauk SSSR,} \textbf{183}, 806 (1968).
%
%
%
\bibitem{wkuz70}
K.  Walasek and A. L. Kuzemsky, Kinetic equations for a system weakly coupled to a thermal bath.
{\em Teor. Mat. Fiz.,}  \textbf{4}, 267 (1970).
%
%
\bibitem{koz08}
V. V. Kozlov,
Gibbs ensembles, equidistribution of the energy of
sympathetic oscillators and statistical models of thermostat.
 \emph{Regular and Chaotic Dynamics}  \textbf{13}, no.~3, 141 (2008).
%
%
%
\bibitem{kuzem06}
A. L. Kuzemsky,  Statistical theory of spin relaxation and diffusion in solids.
\emph{J. Low Temp. Phys.}  \textbf{143}, 213 (2006).    
%
%
\bibitem{wzkuz70}
K.  Walasek, D. N. Zubarev,  A. L. Kuzemsky, 
Schr\"{o}dinger-type equation with damping for a dynamical system in a thermal 
bath. {\em Teor. Mat. Fiz.}  \textbf{5}, 280 (1970).
%
%
%
\bibitem{kuzpaw72}  A. L. Kuzemsky, A. Pawlikowski. Note on the diagonalization of a quadratic linear form defined on
the set of second quantized fermion operators. \emph{Rep. Math. Phys.}  \textbf{3}, 201  (1972).
%
%
\bibitem{kuzem08}
A. L. Kuzemsky, Works of D. I. Blokhintsev and development of quantum physics.
\emph{Physics of  Particles and Nuclei}  \textbf{39}, no.~2, 137-172  (2008).  
%
%
%
\bibitem{lnc71}
A. L. Kuzemsky and K.  Walasek, On the calculation of the natural width of spectral lines of atom
by the methods of nonequilibrium statistical mechanics. {\it Lett. Nuovo Cimento,}  \textbf{2}, 953 (1971).
%
%
%
\bibitem{kuzem12}
A. L. Kuzemsky,
Generalized Van Hove formula for scattering of neutrons by the nonequilibrium statistical medium.
\emph{Int. J. Mod. Phys.,}   B \textbf{26}, 1250092 (34 pages) (2012).
%

%
%
%
\end{thebibliography}
\end{document}